\documentclass[aps,twocolumn,floatfix]{revtex4-1}
\usepackage{hyperref}
\hypersetup{
    bookmarks=true,         
    unicode=false,          
    pdftoolbar=true,        
    pdfmenubar=true,        
    pdffitwindow=false,     
    pdftitle={Certificate},    
    pdfauthor={Dr. Mehul Makwana},     
    pdfsubject={},   
    pdfcreator={Dr. Mehul Makwana},   
    pdfproducer={},  
    pdfkeywords={Energy-Splitter,} {Networks} {VHE}, 
    pdfnewwindow=true,      
    colorlinks=false,       
    linkcolor=red,          
    citecolor=green,        
    filecolor=magenta,      
    urlcolor=cyan           
    \pdftrailerid{} 
\pdfsuppressptexinfo15 
}
\usepackage{amssymb}
\usepackage{graphicx}
\usepackage{caption}
\usepackage{sidecap}

\usepackage{multirow}
\usepackage[table,xcdraw]{xcolor}
\usepackage{soul}
\usepackage{kantlipsum}
\usepackage{amsmath}
\usepackage{amsfonts}
\usepackage{amssymb}
\usepackage{dcolumn} 
\usepackage{braket}
\usepackage{booktabs}
\usepackage{multirow}
\usepackage[table]{xcolor}
\usepackage{dsfont}
\usepackage{times}
\usepackage{epsfig}
\usepackage{xcolor}
\usepackage{tikz}
\usepackage{float}
\usepackage{color}
\usepackage{inputenc}
\usepackage{booktabs}
\usepackage{pgfplotstable}
\usepackage{caption,subcaption}
\def\bx {{\bf x}}

\def\bkappa {\boldsymbol{\kappa}}

\newcommand{\beq}{\begin{equation}}
\newcommand{\eeq}{\end{equation}}
\newcommand{\ba}{\begin{eqnarray}}
\newcommand{\ea}{\end{eqnarray}}

\input epsf

\usepackage[english]{babel}

\begin{document}

\title[]{\hspace{2cm} Designing multi-directional energy-splitters and \newline topological valley supernetworks}
\author{Mehul P. Makwana$^{1, 2}$ and Richard V. Craster$^{1}$}
\affiliation{$^1$ Department of Mathematics, Imperial College London, London SW7 2AZ, UK }
\affiliation{$^2$ Multiwave Technologies AG, 3 Chemin du Pr\^{e} Fleuri, 1228, Geneva, Switzerland}

\begin{abstract}
Using group theoretic and topological concepts, together with tunneling phenomena, we geometrically design interfacial wave networks that contain splitters which partition energy in 2, 3, 4 or 5 directions. This enriches the valleytronics literature that has, so far, been limited to 2-directional splitters. Additionally, we describe a design paradigm that gives greater detail, about the relative transmission along outgoing leads, away from a junction; previously only the negligible transmission leads were predictable. We utilise semi-analytic numerical simulations, as opposed to finite element methods, to clearly illustrate all of these features with highly resolved edge states. As a consequence of this theory, novel networks, with directionality tunable by geometry, ideal for applications such as beam-splitters, switches and filters are created. Coupling these novel networks, that contain multi-directional energy-splitters, culminates in the first realization of a topological supernetwork. 

\end{abstract}
\maketitle
\section{Introduction}
A fundamental understanding of the manipulation and channeling of wave
energy underpins advances in electronic properties, acoustic switches,
optical devices, vibration control and electromagnetism. Guiding waves, splitting and redirecting them between
channels, and steering waves around sharp bends, in a robust and lossless manner
is of interest across many areas of engineering and physics
\cite{mekis_high_1996, yariv_coupled-resonator_1999, chutinan_wider_2002, qiao_current_2014-1, ju_topological_2015, liu_multimode_2004, ma_guiding_2015}. Recent advances based upon ideas originating from
topological insulators \cite{kane_z2_2005, ren_single-valley_2015-1, xiao_valley-contrasting_2007-1, khanikaev_two-dimensional_2017-1}, translated to Newtonian
wave systems, have inspired great interest: In particular, geometrically engineering  topological
photonic and phononic crystals \cite{lu_dirac_2014-1,
  khanikaev_two-dimensional_2017-1} to 
direct waves along interfaces in a robust tuneable manner 
 has shown much potential.  In this article, we leverage the efforts by the 
topological valleytronics community \cite{gao_valley_2017, dong_valley_2017-1, yang_topological_2018, kang_pseudo-spinvalley_2018, chen_tunable_2018, he_silicon--insulator_2018, lu_observation_2017-1, ye_observation_2017, zhang_topological_2018, wu_direct_2017, jung_active_2018, gao_topologically_2017, zhang_manipulation_2018, xia_observation_2018,  shalaev_experimental_2017, liu_tunable_2018}, to 
design a range of novel interfacial wave networks. These extend the 
interfacial network designs prevalent in the current literature by allowing for more than 
 a 2-way splitting of energy away from a nodal region. For  hexagonal structures there are three distinct edges which yield up to three sets of edge states \cite{makwana_geometrically_2018, lu_dirac_2014-1}; the current literature has concentrated upon only one of these. Here we analyze the remaining two, one of which is topological, whilst the other is not. Despite that latter state being non-topological, the large separation in Fourier space between opposite propagating modes results in an interfacial wave that is relatively robust to sharp disorders.
 To elucidate our principles with clarity we outline a comprehensive design paradigm which is in turn utilised to build novel networks.

 \begin{figure}[h!]
 \centering
 \begin{tabular}{llllll}
 \includegraphics[scale=0.155]{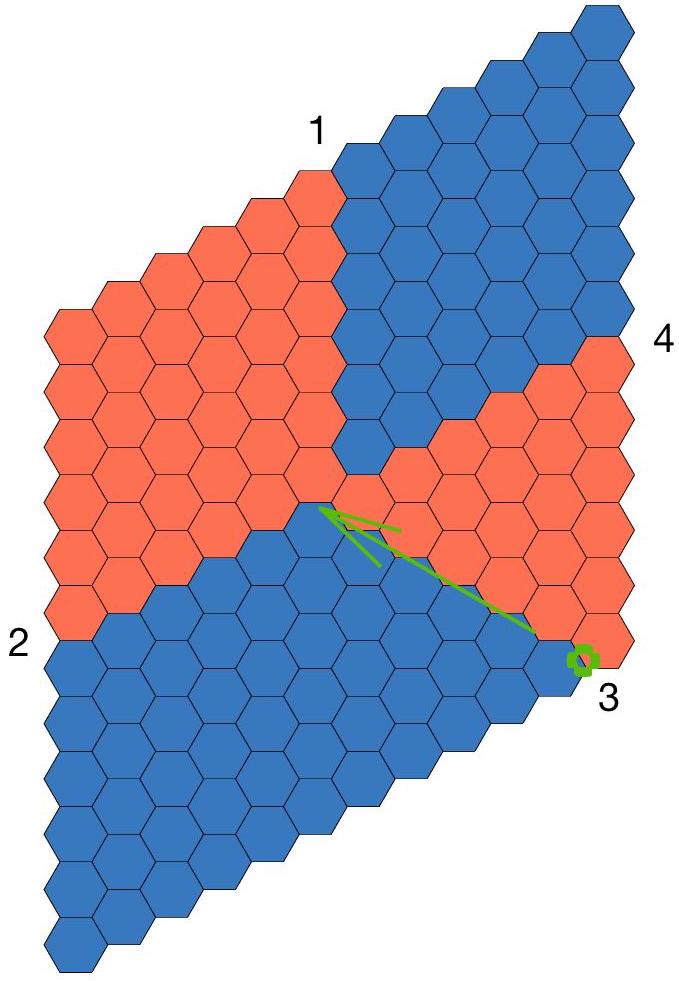} &
 (a)	\includegraphics[scale=0.260]{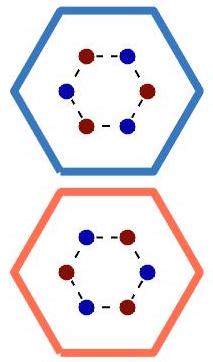} &
 (b)   \includegraphics[scale=0.1275]{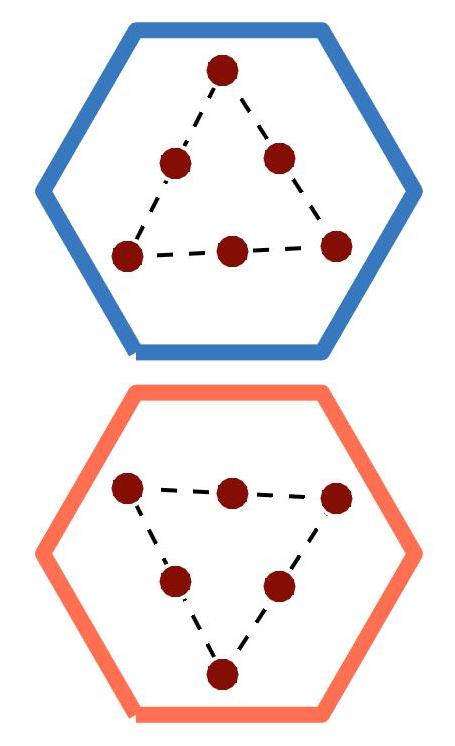}
 \end{tabular}
 
 \centering
(c) \includegraphics[scale=0.13]{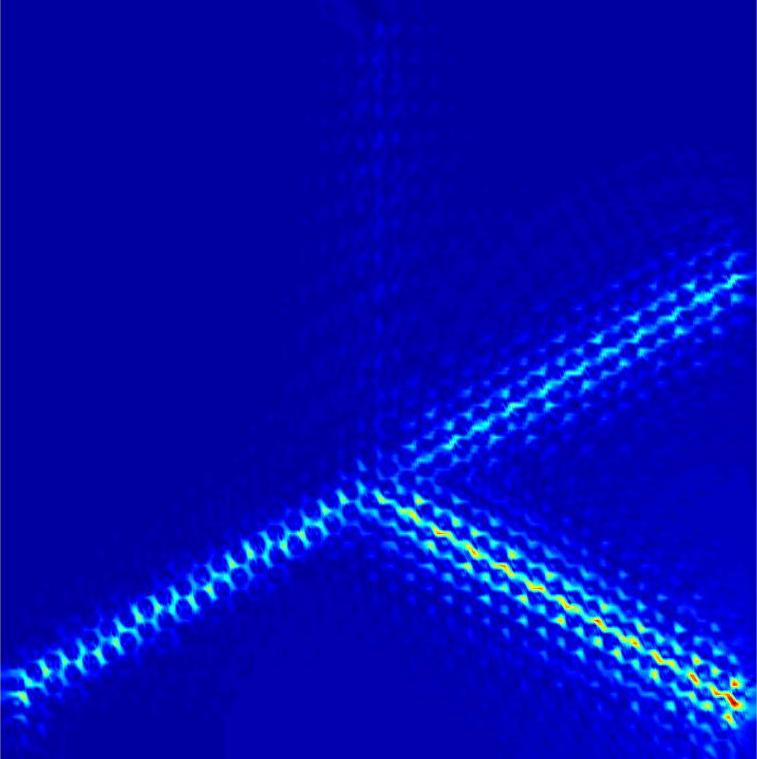}
(d)	\includegraphics[scale=0.140]{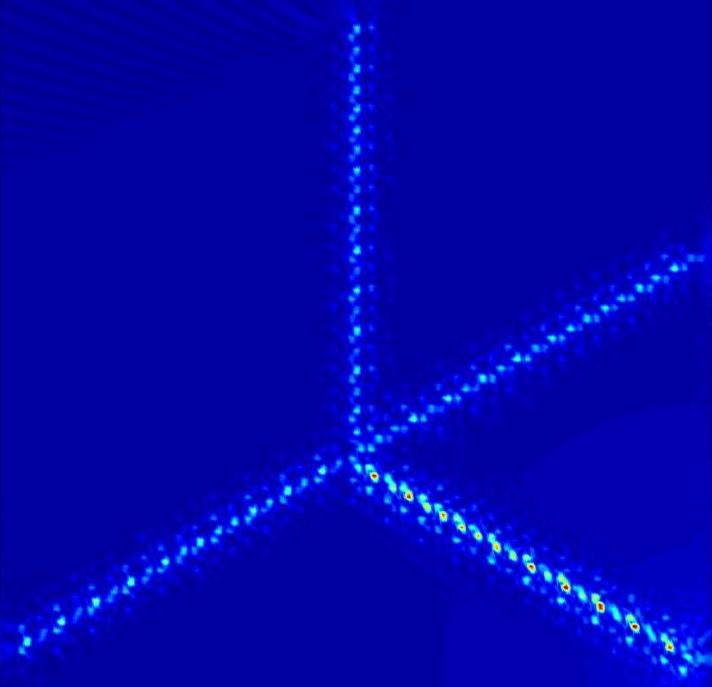}
 \caption{Intelligently constructed domain comprised of geometrically distinct regions; mass-loaded structured elastic plate is the model chosen. Source is placed at the start of the interface 3. Arrangement of masses for blue and orange cells are shown in panels (a) and (b); specific system parameters are detailed in captions of Figs. \ref{fig:C6v_DC} and \ref{fig:C3v_DC}. Panels (a) and (b) are associated with the (c) and (d) scattered field panels. Panel (c) shows typical 2-way energy-splitting \cite{cheng_acoustic_2016}; an alternative geometrically engineered interfacial wave network, with more than 2-way splitting, is shown in (d).}
\label{fig:NatMat_Ex}
 \end{figure}

Recent attempts to leverage the properties of quantum topological
effects to design, so-called, topological power-splitters, for
continuous Newtonian systems \cite{cheng_robust_2016-1,cheng_acoustic_2016, xiaoxiao_direct_2017, xia_topological_2017-1}
would benefit from a clear design paradigm 
 explaining how to partition the energy of
topological modes. 
 Splitters, and efficient transport
around sharp bends, are often achieved using a different mechanism,
that of cavity waveguides in photonic crystals \cite{mekis_high_1996, yariv_coupled-resonator_1999, liu_multimode_2004, chutinan_wider_2002}. Given that we are dealing with interfacial waves,
the power-splitting mechanism, espoused herein 
 is an alternative means to split energy to those found in \cite{yariv_coupled-resonator_1999, liu_multimode_2004}.

By focussing upon the underlying concepts of time-reversal symmetric (TRS) valley-Hall 
 insulators \cite{pal_edge_2017, ma_all-si_2016, ren_topological_2016-1, lu_valley_2016-1, lu_observation_2017-1, chen_valley-contrasting_2017-1, xiaoxiao_direct_2017}, 
 that  
%
 do not break TRS, 
 our system is ultimately topologically
 trivial. Despite this, valley-Hall insulators do have  advantages;
 they are relatively straightforward to design as we only need to break
spatial inversion and/or a reflection symmetry, together with 
 proactively suppressing backscattering between modes of opposite group velocity.  

The group theoretic and topological concepts foundational to our
approach hold irrespective of any specific two-dimensional scalar
wave system. 
 We illustrate these concepts using a single system, specifically, a
 structured thin elastic Kirchhoff-Love (K-L) plate
\cite{landau70a} 
 for which
  many results for point scatterers are explicitly
available \cite{evans07a}; the ideas themselves carry across to photonics,
phononics and plasmonics. 
 A particularly pleasant feature of the
K-L model is that the fundamental Green's function is,
unlike acoustics and electromagnetism, non-singular and bounded
thereby 
 simplifying simulations.  
 
We begin by briefly formulating the Bloch eigenstate and scattering problems in the context of the K-L elastic plate, Sec.  \ref{sec:formulation}, and then move on, Sec. \ref{sec:three}, to describe the construction, origin and classification of the three canonical edge states that are possible. Sec. \ref{sec:network_paradigm} introduces the design paradigm for creating networks and we elucidate the critical points required in building or interpreting networks: sharp modal shapes, filtering, Fourier space separation between opposite propagating modes, chirality and phase matching, tunneling of energy and the effect of the nodal region at the junction between interfaces. Given the paradigm developed we move on, to Sec. \ref{sec:building}, where we construct such novel interfacial wave networks; we demonstrate the collective strength of the design principles, in Sec. \ref{sec:supernetwork}, by building a large scale topological supernetwork. Finally, in Sec. \ref{sec:concluding}, we pull together concluding remarks.

\subsection{Formulation}
\label{sec:formulation}
Displacement Bloch eigenstates  $\psi_{n
  \bkappa}({\bf x})$ satisfy the 
(non-dimensionalized) K-L equation
\begin{equation}
\left[\nabla_{\bf x}^4 
 -\omega_{\bkappa}^2\right]\psi_{j \bkappa}=F(\bx),
\label{eq:kirchhoff}
\end{equation} 
 for Bloch-wavevector ${\bkappa}$, $j$ labelling the eigenmodes and
 $\omega_{\bkappa}$ the frequency; 
 reaction forces at the point constraints, $F(\bx)$, introduce
 dependence upon the direct lattice. 

In two-dimensional systems there are only three symmetry sets that 
 yield Dirac cones \cite{makwana_geometrically_2018, lu_dirac_2014-1} of which we use two. 
 The gapping of these Dirac cones is done via two distinct symmetry-breaking mechanisms; for point scatterers this entails varying their masses or positions. 
The simplest model to use is that of the mass-loaded elastic plate
where  
the reaction forces are proportional to the displacement and hence,
\beq
F(\bx)=\omega_{\bkappa}^2\sum_{\bf n}\sum_{p=1}^{P}
 M^{(p)}_{\bf n} \psi_{j \bkappa}(\bx)\delta\left({\bf x}-{\bf x}^{(p)}_{\bf n}\right).
\label{eq:ML_RHS}
\eeq 
Here ${\bf n}$ labels each elementary cell,  containing 
 $p=1...P$ constraints, that repeats 
periodically. Eq. (\ref{eq:kirchhoff}) is
solved to obtain the eigenstates using 
plane wave expansions \cite{johnson01a}, modified for elastic
plates \cite{xiao12a}, and when forcing is applied  we utilize a
Green's function approach \cite{torrent13a} where the total wavefield is
given for $N$ scatterers by
\beq
\psi_{j\bkappa}({\bf x}) = \psi_{i}({\bf x}) + \sum_{\bf n}\sum_{p=1}^{P} F_{\bf n}^{(p)} g\left(\omega_{\bkappa}, |{\bf x}-{\bf x}_{\bf n}^{(p)}|\right)
\eeq 
with $\psi_i$ is the incident field. 
Using the well-known Green's function \cite{evans07a}, \( 
g\left(\omega_{\bkappa}, \rho\right) = ({i}/{8\omega_{\bkappa}^2})
\left[H_0(\omega_{\bkappa} \rho) - H_0(i\omega_{\bkappa} \rho)\right]
\), the unknown reaction terms $F_{\bf n}^{(p)}$ come from the linear system
\begin{multline}
F_{\bf n}^{(p)} = 
M_{\bf n}^{(p)}\omega_{\bkappa}^2\Big[\psi_{i}\left({\bf x}_{\bf n}^{(p)}\right) + \\ 
\sum_{\bf m}\sum_{q=1}^{P}
  F_{\bf m}^{(q)} g\left(\omega_{\bkappa}, |{\bf x}_{\bf m}^{(q)}-{\bf x}_{\bf n}^{(p)}|\right) \bigg].
  \label{eq:linear_system}
\end{multline}
This model has considerable advantages in terms of being almost completely explicit, and additionally the Green's function is non-singular; this leads to highly resolved solutions and edge states that enable us to interpret the results accurately. The numerical schemes that emerge from this approach are efficient thereby allowing us to concentrate on the design process itself.   

 \begin{figure}[ht!]
 \centering
 \begin{tabular}{llll}
(a) 	\includegraphics[scale=0.205]{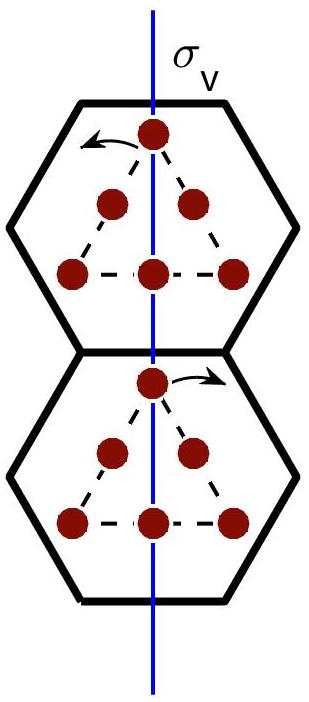} &
 (b)	\includegraphics[scale=0.205]{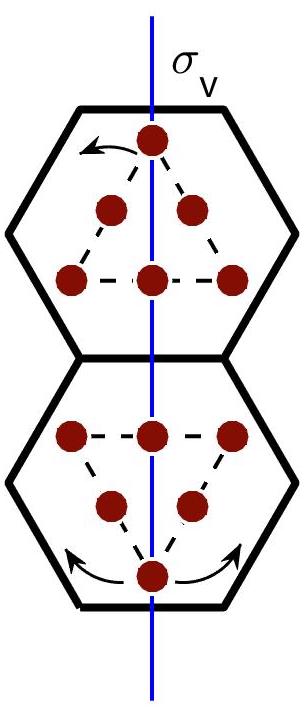}	
\end{tabular}

 \caption{The geometric creation of interfacial edge states relies upon broken six-fold symmetry; the stacked media are required to share the same band-gap, the latter arises from having same angular perturbation away from the reflection line, $\sigma_v$. The details of the 3 edges are summarised in Table 1. Perturbations shown in (a) lead to a type 1 edge, resulting in similar edge modes to those of Fig. \ref{fig:C6v_DC}; whilst the perturbations indicated by the left and right arrows, for the lower cell in (b), pertain to 
the type 2 (Fig. \ref{fig:C3v_DC}) and type 3  (Fig. \ref{fig:C3v_nonorthog_DC}) edges, respectively.}
\label{fig:edges}
 \end{figure}

\subsection{Three distinct edges: Topology, symmetry and the cellular structure}
\label{sec:three}

\onecolumngrid

\begin{table}[t!]
\begin{tabular}{|
>{\columncolor[HTML]{FFFFFF}}l |c|c|c|c|c|}
\hline
\multicolumn{1}{|c|}{\cellcolor[HTML]{FFFFFF}\textbf{Edge Type}} & \cellcolor[HTML]{FFFFFF}\textbf{\begin{tabular}[c]{@{}c@{}}Point Group Symmetries\\ pre-perturbation\end{tabular}} & \cellcolor[HTML]{FFFFFF}                                                                                                 & \cellcolor[HTML]{FFFFFF}                                                                                                & \cellcolor[HTML]{FFFFFF}                                                                                            &                                         \\
\multicolumn{1}{|c|}{\cellcolor[HTML]{FFFFFF}$\rightarrow$}      & \cellcolor[HTML]{FFFFFF}${G_{\Gamma}, G_{K,K'}}$                                                                   & \multirow{-2}{*}{\cellcolor[HTML]{FFFFFF}\textbf{\begin{tabular}[c]{@{}c@{}}Medium 1 \\ post-perturbation\end{tabular}}} & \multirow{-2}{*}{\cellcolor[HTML]{FFFFFF}\textbf{\begin{tabular}[c]{@{}c@{}}Medium 2\\ post-perturbation\end{tabular}}} & \multirow{-2}{*}{\cellcolor[HTML]{FFFFFF}\textbf{\begin{tabular}[c]{@{}c@{}}Topological\\ Protection\end{tabular}}} & \multirow{-2}{*}{\textbf{System abbreviation}} \\ \hline
\cellcolor[HTML]{FFFFFF}                                         &                                                                                                                    & $M_1 = M_0, $                                                                                                            & $M_1 = M_0 + \beta,$                                                                                                    &                                                                                                                     &                                         \\
\multirow{-2}{*}{\cellcolor[HTML]{FFFFFF}\textit{Type 1}}        & \multirow{-2}{*}{${C_{6v}, C_{3v}}$}                                                                               & $ M_2 = M_0 + \beta$                                                                                                     & $ M_2 = M_0 $                                                                                                           &                                                                                                                     & \multirow{-2}{*}{$C_{6v}$ nontrivial}          \\ \cline{2-4} \cline{6-6} 
concave and convex curves                                        & ${C_{3v}, C_{3v}}$                                                                                                 & $ \theta = + \alpha$                                                                                                     & $\theta = - \alpha$                                                                                                     & \multirow{-3}{*}{Yes}                                                                                               & ---                                   \\ \hline
\textit{Type 2}                                                  &                                                                                                                    &                                                                                                                          &                                                                                                                         &                                                                                                                     &                                         \\
one convex curve                                                 & \multirow{-2}{*}{${C_{3v}, C_{3v}}$}                                                                               & \multirow{-2}{*}{$ \theta = + \alpha$}                                                                                   & \multirow{-2}{*}{$ \theta = + \alpha + \pi/3$}                                                                          & \multirow{-2}{*}{Yes}                                                                                               & \multirow{-2}{*}{$ C_{3v}$ nontrivial}         \\ \hline
\textit{Type 3}                                                  &                                                                                                                    &                                                                                                                          &                                                                                                                         &                                                                                                                     &                                         \\
two convex curves                                                & \multirow{-2}{*}{${C_{3v}, C_{3v}}$}                                                                               & \multirow{-2}{*}{$ \theta = + \alpha$}                                                                                   & \multirow{-2}{*}{$ \theta = - \alpha + \pi/3$}                                                                          & \multirow{-2}{*}{No}                                                                                                & \multirow{-2}{*}{$ C_{3v}$ trivial}          \\ \hline
\end{tabular}
\caption{Summary of the three edge types leading to Figs. \ref{fig:C6v_DC}, \ref{fig:C3v_DC} and \ref{fig:C3v_nonorthog_DC}. In the  $C_{6v}$ case, $M_1$ and $M_2$ denote alternating mass values 
within a hexagon (see Fig. \ref{fig:C6v_DC}); $M_0$ denotes the unperturbed mass value and $\beta$ the perturbation. For the $C_{3v}$ cases, $\theta$ represents the  angular perturbation, $\alpha$, away from the reflection line $\sigma_v$ (see Fig. \ref{fig:edges}). Topological protection follows when the valley Chern numbers are opposite for adjoining media. Abbreviations are adopted to concisely distinguish between the three systems referenced throughout this article.}
\label{3edges}
\end{table}

\twocolumngrid

We demonstrate three distinct edge states that are intelligently constructed; two of these are topologically nontrivial whilst one is topologically trivial. Despite the latter case being trivial it will be shown in Sec. \ref{sec:Fourier_space}, that it is still relatively robust to backscattering due to the Fourier space separation between the forward and backward propagating modes; typically the well-studied topological trivial interfacial and cavity guide states effectively  rely solely upon this separation. A summary of the three types is shown in table I and their geometrical origins are visually demonstrated in Fig. \ref{fig:edges}.

 Turning our attention toward the topological nontrivial states, 
 the valley Hall effect originates from the 
gapping of Dirac cones resulting in 
nontrivial band gaps where broadband edge states are guaranteed to reside; 
simply placing two media, that share a band gap, as neighbours does
not guarantee an interfacial mode \cite{delplace_zak_2011}. The topological invariant
that dictates the construction of our neighbouring media is the valley Chern number;
this takes non-zero values locally at the $KK'$ valleys. By attaching two media, with opposite valley Chern numbers, broadband chiral edge states arise; these interface states are commonly known as topological confinement states, kink states, zero modes, or zero-line modes (ZLMs). From hereon in we use the term ZLM to refer to these topologically nontrivial states; the etymology of this term arises from the adjoining media, either side of the interface, having opposing valley Chern numbers.

We generate ZLMs (and incidentally also the topologically nontrivial modes) by 
placing one gapped medium above another; this second medium could either be a reflection and/or $\pi/3$ rotation of the first. The simplicity
of this construction, and the apriori knowledge of how to tessellate the
two media, to produce these broadband edge states, is the main
benefit of these geometrically engineered modes. A benefit of the topologically nontrivial valley modes is that the opposing valley Chern numbers imbue the edge states with an additional protective property (Sec. \ref{sec:chirality_phase}); despite the type 3 edge (table \ref{3edges}) yielding topologically nontrivial states it ultimately shares many of the same features, as its topological counterpart, in a practical setting due to the Fourier space separation of the counter-propagating modes (Sec. \ref{sec:Fourier_space}).

For instance, for the $C_{6v}$ nontrivial case, from \cite{makwana_geometrically_2018},  the effective bulk Hamiltonian takes the form, 
\beq
H_{\text{eff}} = \tau_z v_D(\hat{\sigma}_z \Delta\kappa_x - \hat{\sigma}_x \Delta\kappa_y) + \tau_z
\mathcal{M}_{K}\hat{\sigma}_y,
\label{eq:pert_Ham_term}
\eeq 
where $\mathcal{M}_{K} = \omega_{K}^2 \Delta M, v_D$ is the system dependent group velocity, $\{\hat{\sigma}_i\}$ are the Pauli matrices.
The $\Delta M$ term is responsible for gapping the Dirac point and
differs depending upon the manner whereby 
inversion symmetry is broken; for the canonical honeycomb case, $\Delta M = \beta/2$.
The presence of the valley Pauli matrix, $\tau_z$, relates the Dirac masses at the $KK'$ valleys by $\mathcal{M}_K =-\mathcal{M}_{K'}$. 
The corresponding eigenvalues for this effective Hamiltonian are,
\beq
(\omega_{K}^2-\omega_{\bkappa}^2) = \pm \sqrt{v^2_D |\Delta\bkappa|^2 + \mathcal{M}^2_{K}}.
\label{eq:perturbed_eigenvalues}
\eeq 
This is the form of the eigenvalues for the massive Dirac
fermionic equation albeit for a platonic crystal. Following similar arguments to \cite{ochiai_photonic_2012}
, we evaluate the Chern number as $C
= C_K + C_{K'}$ where $C_{K, K'} = \text{sgn}\left(\mathcal{M}_{K, K'}\right)/2$. 
The term $\mathcal{M}_{K}$ is responsible for gapping the Dirac cones 
by reducing the symmetries of the cellular structures. The
Dirac cone itself is geometrically obtained in three distinct ways, \cite{makwana_geometrically_2018};
these are described by the space group symmetries $C_{6v}, C_{3v}, C_6$ however, for simplicity, in this article we
solely concentrate upon the $C_{6v}, C_{3v}$ cases. We could have used $C_{3v}$ case throughout, i.e. for the type 1 edge, but we opted to include a $C_{6v}$ case to illustrate the generality of our arguments. 

The systematic reduction of these 
cellular structures takes the space group symmetry down to $C_{3}$ and 
consequently reduces the point group symmetries at the $KK'$ valleys to $C_3$;
this reduction at the valleys gaps the Dirac cone. The symmetry reduction down to 
$3$-fold symmetry leads to three symmetrically distinct edges for
each cellular structure.

For the tight-binding
model, 
\cite{bi_role_2015}, ZLMs with distinguishable valley degrees of freedom
exist for every propagation angle except for the armchair;
the armchair termination exactly superposes the $KK'$ valleys thereby
coupling them. In
principle one could use other edge terminations for continua however it is impractical to use fractional
cells, as in \cite{bi_role_2015}, when partitioning the media. Therefore the topological networks that we create are based solely upon the 
zigzag interface as they offer the greatest protection against backscattering.

\begin{figure} [h!]
\centering
	\includegraphics[scale=0.405]{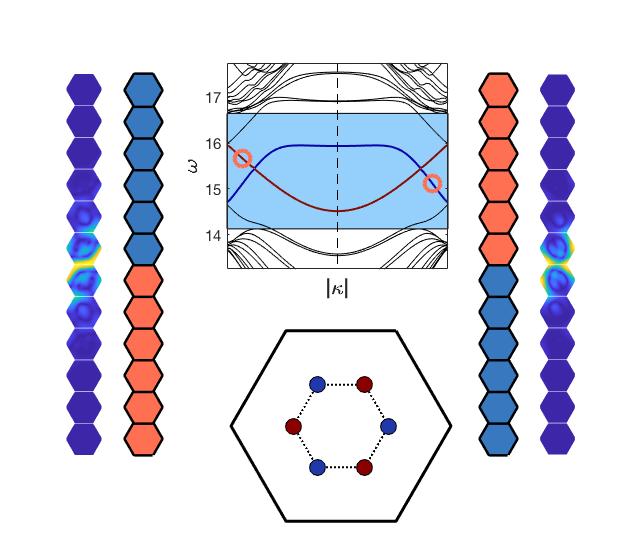}
\caption{The gapped Dirac ZLM (related to type 1 edge of Table \ref{3edges}) 
  with original space group symmetry $C_{6v}$ and alternating the masses $M_{1}=1, M_2 = 2$, which have 
distance from centroid to masses $0.5$ (the pitch is $2$). Left-hand circle on concave curve at
   $\omega = 15.11$ corresponds to the ZLM (left). Right-hand circle on convex curve at
   $\omega = 15.67$ to the ZLM (right). Note, the detail of both of the easily distinguishable edge states; hence
   we can easily attribute a modal pattern to a specific ordering of the adjoining media.
}
\label{fig:C6v_DC}
\end{figure}

\begin{figure} [h!]
\centering
\includegraphics[scale=0.405]{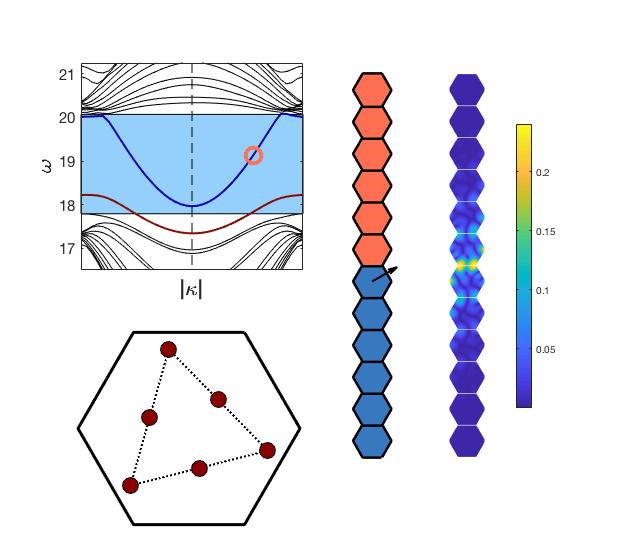}
\caption{The gapped Dirac ZLM 
  with original space group symmetry $C_{3v}$ emerges from the type 2 edge (Table \ref{3edges}). The cellular structure for the upper medium is shown; the lower medium is a $\pi/3$ rotation of this (see Fig. \ref{fig:4cell}). 
  This case has the distance from centroid to vertices of triangle $ = 0.85$, unit masses (the pitch is $2$) and a perturbation of $0.05$.
The circled point at $\omega = 19.13$ corresponds to the ZLM shown. In contrast to Fig. \ref{fig:C6v_DC} we have a broad frequency range for which there is a non-simultaneous edge mode. The interface for the broadband edge mode is explicitly shown in Fig. \ref{fig:4cell}(a), the narrowband zigzag edge is shown in Fig. \ref{fig:4cell}(b). The lack of overlap between the states will be utilized in  Secs. \ref{sec:filtering}, \ref{sec:filter_network} when we wish to preferentiate the energy propagation along particular leads within a network. For a type 2 edge, you are not guaranteed a non-simultaneous edge mode.
 It is imperative to analyze the three distinct types of edges (see Fig. \ref{fig:edges} and Table \ref{3edges}) in order to discern whether they produce different edge states and hence different scattering behaviour within a network. The colourbar is shown to emphasise that a graded colour scheme is used for all displacements within this article; often an ungraded scheme is used within the valleytronics community which can be visually misleading.  
  }
\label{fig:C3v_DC}
\end{figure}

 \begin{figure}[h!]
 \centering
 \begin{tabular}{llll}
(a) 	\includegraphics[scale=0.135]{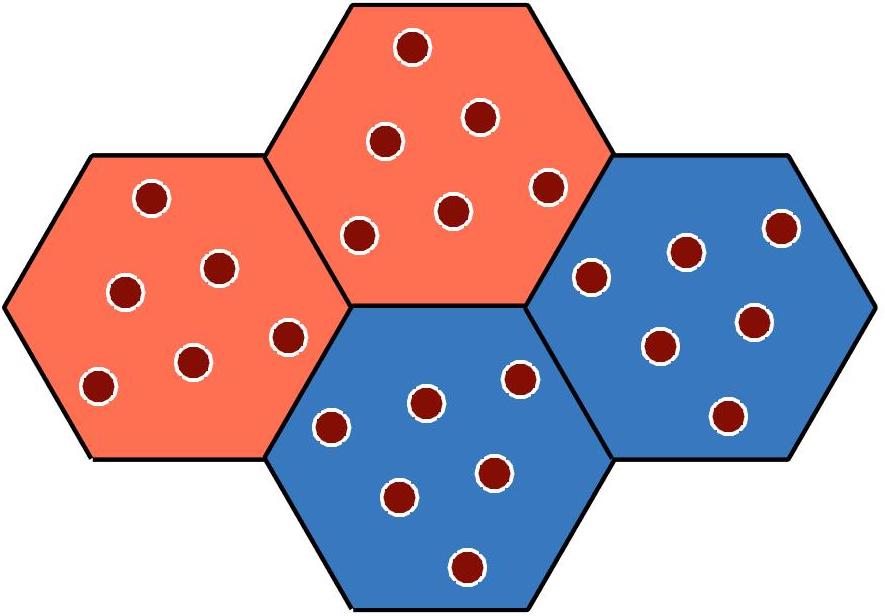}	&
 (b)	\includegraphics[scale=0.145]{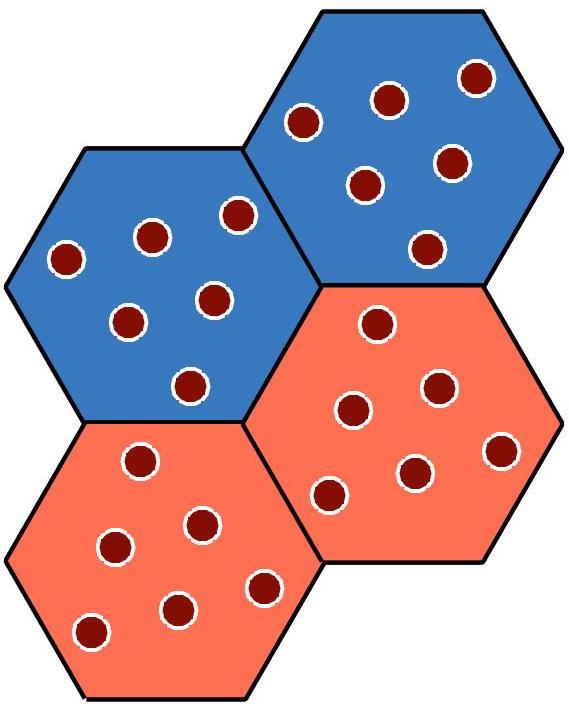}	
\end{tabular}
\caption {The zigzag edges for orange medium over blue and vice-versa for the structure used in Fig. \ref{fig:C3v_DC} and associated with the type 2 edge of Table \ref{3edges}.
}
\label{fig:4cell}

 \end{figure}

 \begin{figure}[h!]
 \centering
 \begin{tabular}{llll}
(a) 	\includegraphics[scale=0.135]{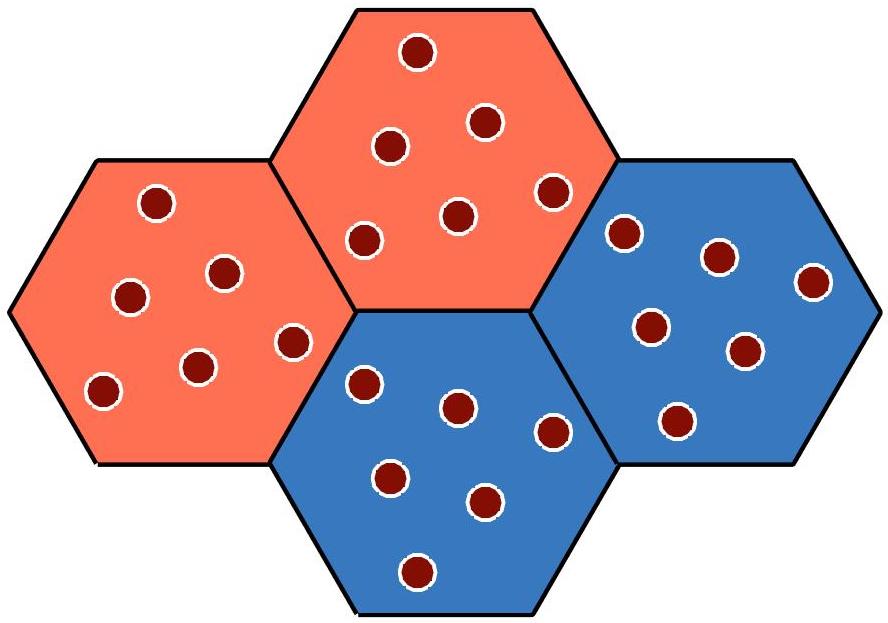}	&
 (b)	\includegraphics[scale=0.145]{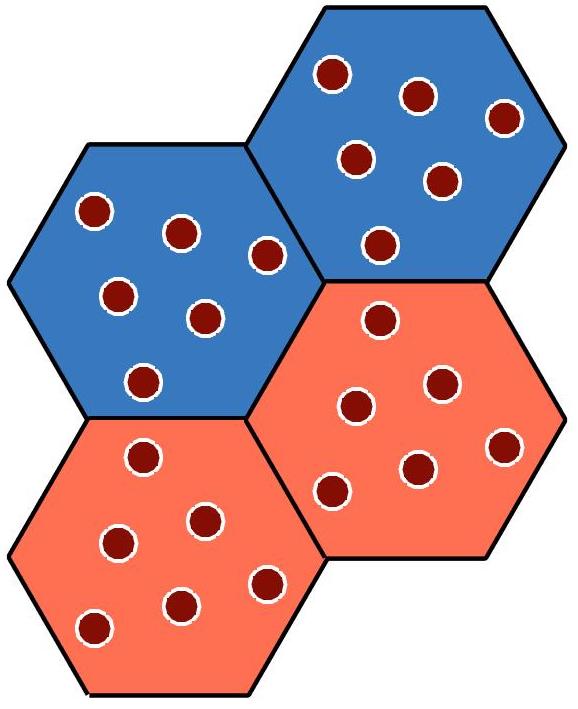}	
\end{tabular}
\caption{The zigzag edges for orange medium over blue and vice-versa used in Fig. \ref{fig:C3v_nonorthog_DC} and associated with the type 3 edge of Table \ref{3edges}. The zigzag edges in panels (a) and (b) are nearly identical which gives the almost overlapping edge modes shown in Fig. \ref{fig:C3v_nonorthog_DC}}
\label{fig:nonorthog_4cell}

 \end{figure}

\begin{figure} [h!]
\centering
 \includegraphics[scale=0.170]{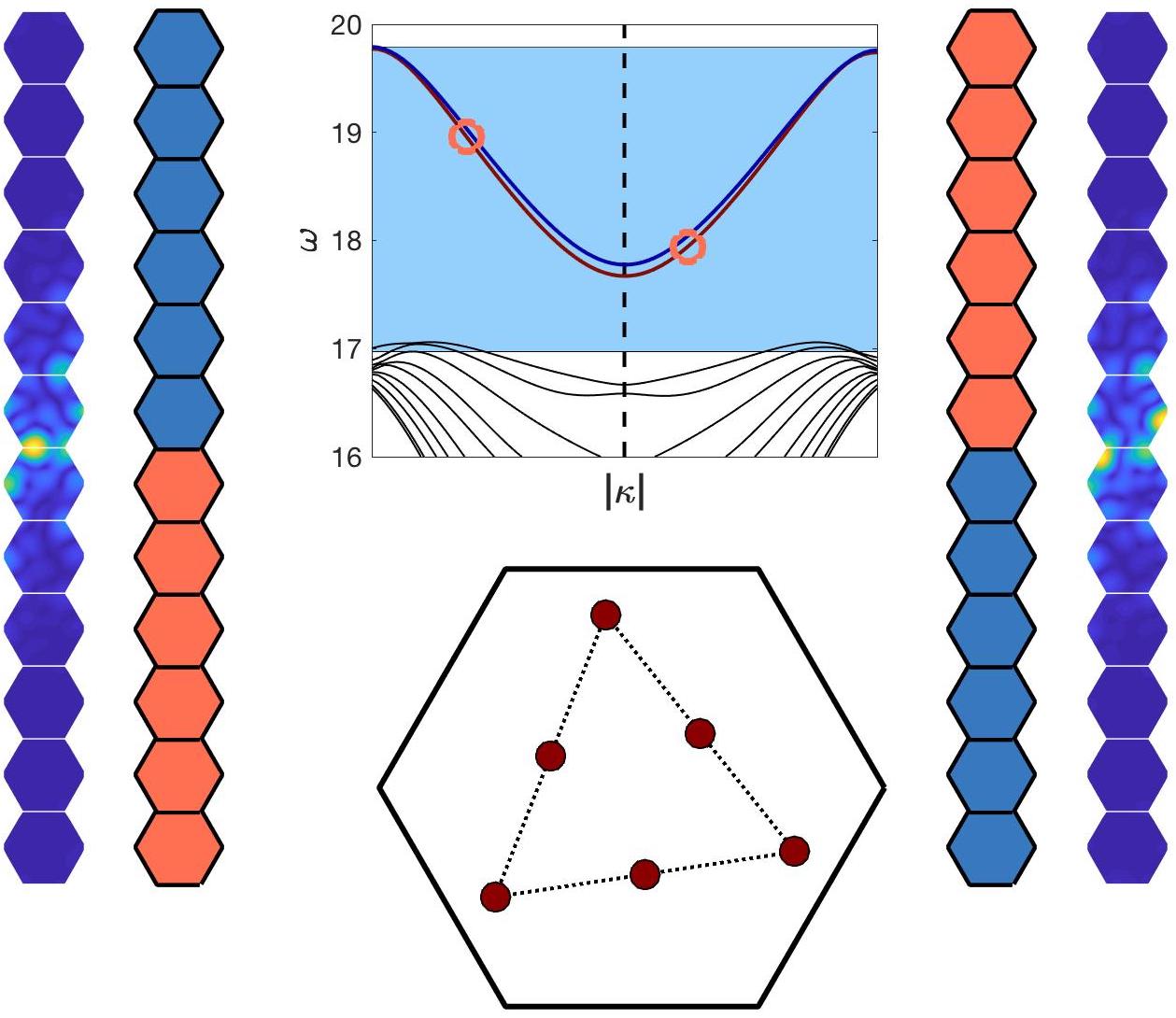}
\caption{The gapped Dirac topologically trivial edge state with original space group symmetry $C_{3v}$, arises from the type 3 edge (Table \ref{3edges}).  The trivial nature is due to the Chern numbers at the $KK'$ valleys being identical; despite this, the simultaneous bulk band-gap, for the two media, and their relative difference in orientation results in broadband edge states. Distance from centroid to vertices of triangle $ = 0.85$ and unit masses (the pitch is $2$). Similar to Fig. \ref{fig:C6v_DC} we obtain simultaneous edge states albeit for the type 3 edge we have two concave curves as opposed to a convex and a concave. Left-hand circle at $\omega = 18.95$ corresponds to the interfacial mode on the left whilst the right-hand circle, $\omega = 17.95$ to the right-hand mode. }  
\label{fig:C3v_nonorthog_DC}
\end{figure}

\begin{figure}[h!]
 \centering
 \begin{tabular}{llll}
(a) 	& \includegraphics[width=3cm]{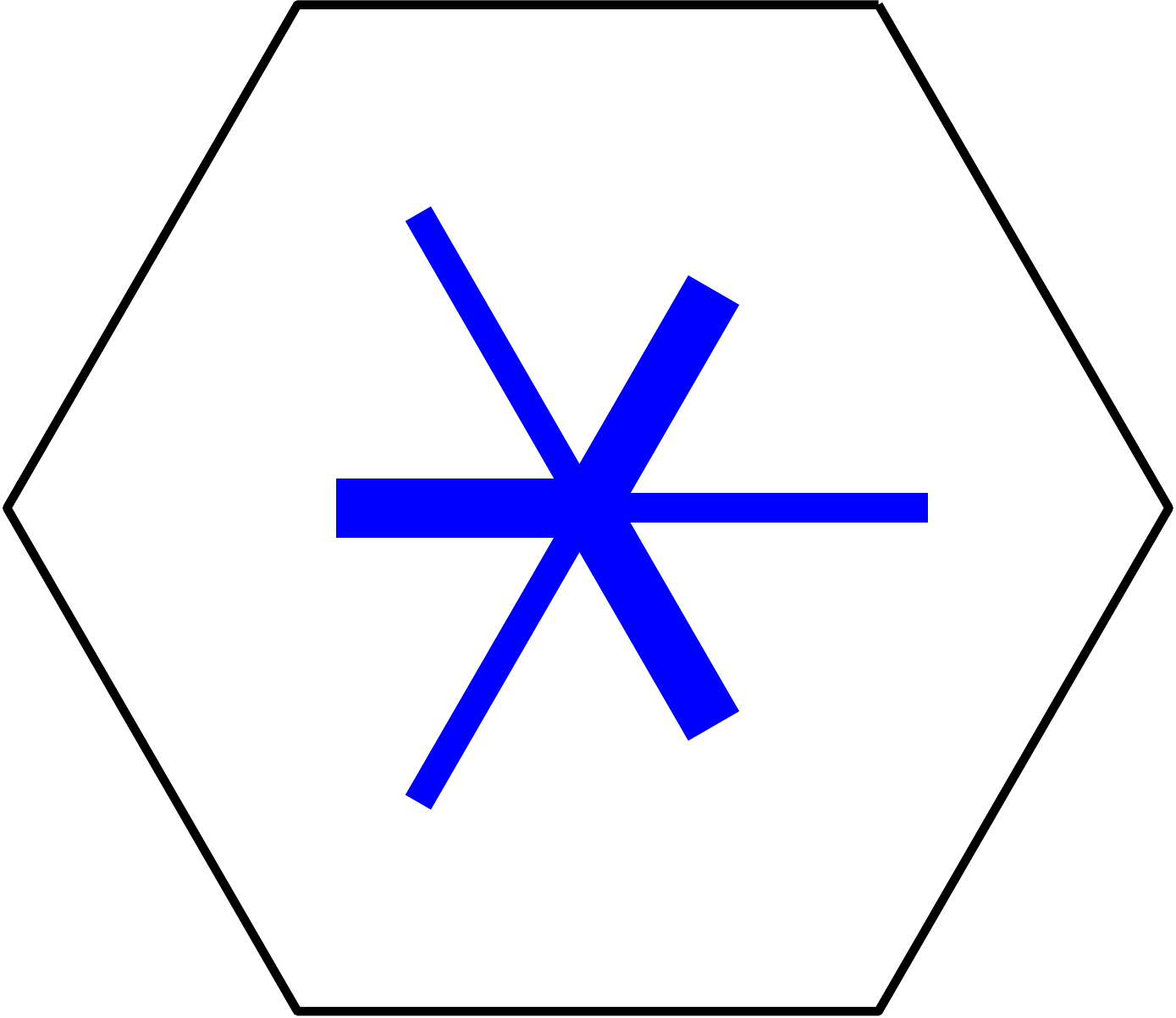}
 (b)		\includegraphics[width=3cm]{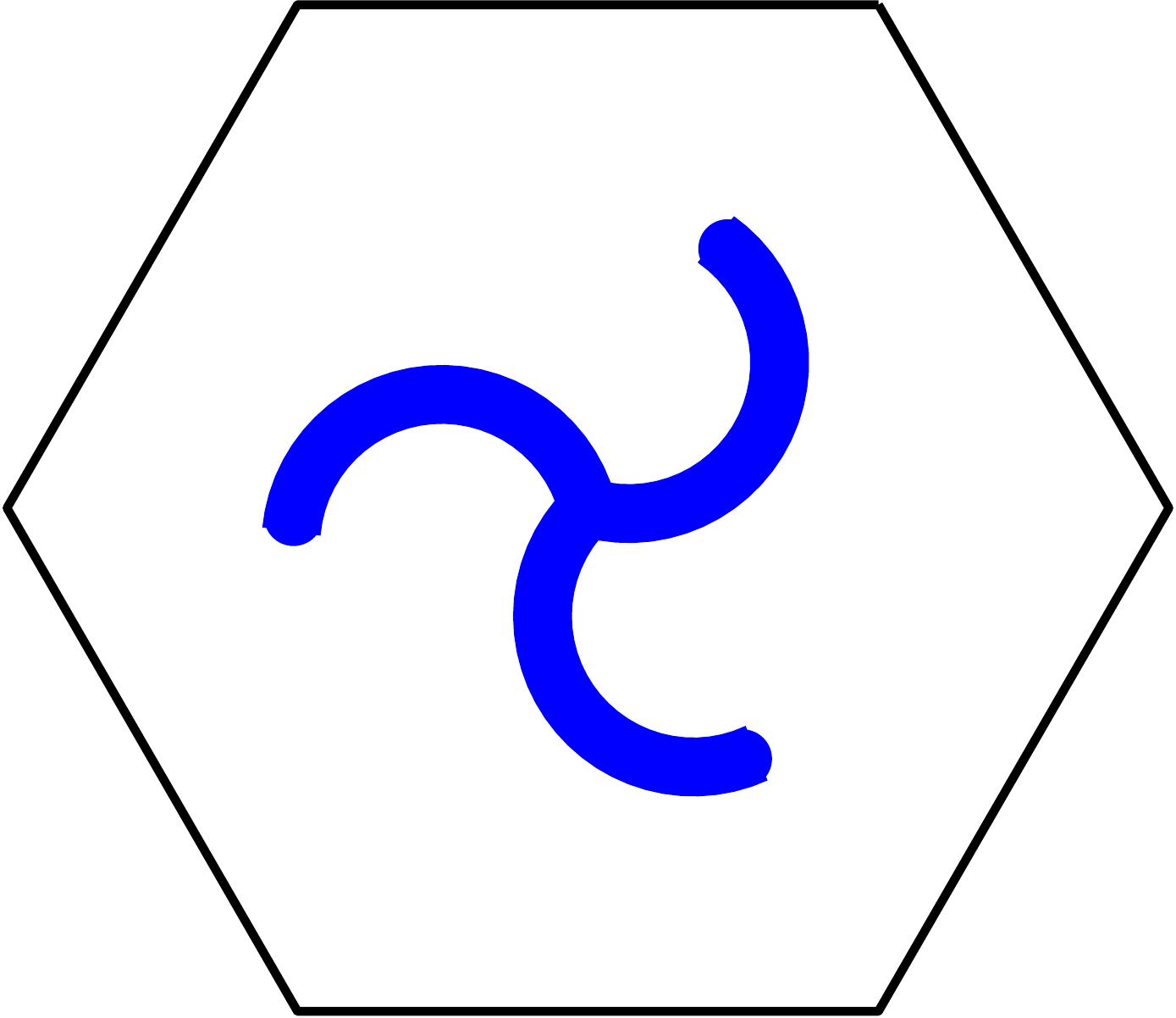}
\end{tabular}
	\caption{Structural elements moving beyond point masses: (a) snowflake: continuous analogue of $C_{6v}$ case, (b) Isle of Man: continuous analogue of $C_{3v}$ case.    }  
\label{fig:other_cells}

 \end{figure}

To summarize, we have identified the canonical three types of edges that arise from breaking six-fold symmetry in hexagonal structures and the resulting edge states as shown in Figs \ref{fig:C6v_DC}, \ref{fig:C3v_DC}, \ref{fig:C3v_nonorthog_DC}. We present a systematic breakdown of the different edge modes that may arise from breaking the 
symmetry induced Dirac cones occurring for the hexagonal lattice and thence for all 2D media. Previously, the bulk of the valleytronics literature \cite{gao_valley_2017, dong_valley_2017-1, yang_topological_2018, kang_pseudo-spinvalley_2018, chen_tunable_2018, he_silicon--insulator_2018, lu_observation_2017-1, ye_observation_2017, zhang_topological_2018, wu_direct_2017, jung_active_2018, gao_topologically_2017, zhang_manipulation_2018, xia_observation_2018, shalaev_experimental_2017, liu_tunable_2018} have exclusively only dealt with the type 1 edge (in the notation of Table \ref{3edges}). We now utilize the unexplored type 2 and 3 edges to demonstrate their properties for controlling and redirecting waves. Note that the simplest $C_{6v}$ case is actually the honeycomb structure; we have opted for hexagonal arrangement here because of its parallels with continuous inclusions (see Fig. \ref{fig:other_cells}), the $C_{3v}$ arrangements can also be mapped over to continuous inclusions.

\section{Network design paradigm}
\label{sec:network_paradigm}
We outline the design principles that will be used in the subsequent section for creating novel topological networks. 
The two nontrivial modes that we are studying are characterized as weak topological states, protected solely by symmetry, hence care must be taken to prohibit backscattering. A set of principles regarding, the optimization of these valley modes, was given in \cite{qian_theory_2018}. To clarify, the protection arises both, from the opposite chirality of opposite propagating modes and the intervalley Fourier separation between these two states. Only the former is a topological effect, whilst the latter also occurs for topologically trivial interfacial and cavity waveguide modes. Hence, for the $C_{6v}$ and $C_{3v}$ nontrivial systems we have both of these protective mechanisms, however for the $C_{3v}$ nontrivial case we solely rely on the latter; despite this in Sec. \ref{sec:Fourier_space} we demonstrate how these modes still appear robust against sharp disorder (i.e. the turning point at the junction) for a broadband range of frequencies. Another benefit of these trivial interfacial modes is that they afford  unrestricted directional splitting as  compared with their nontrivial counterparts (Sec. \ref{sec:chirality_phase}).
\\

Our design paradigm is similar to \cite{qian_theory_2018} albeit our application is slightly different; our aim is to build robust networks, comprised of trivial or nontrivial interfacial modes, not just to characterise the robustness of nontrivial ZLMs. Therefore, in addition to the robustness of the modes, we require additional features to aid the tunability of energy as it propagates within an interfacial wave network. An outline of the six  principles is given below, before we embark upon more detailed numerical explanations, in the subsequent subsections:

\begin{itemize}

\item \emph{Different modal shapes.}\textemdash Our semi-analytic expressions allow us to  obtain precise and sharp modes where the distinct modal patterns are easily seen. We distinguish the modes present, those related to medium 1 over medium 2 and its reverse; by solving the linear system   \eqref{eq:linear_system} we easily visualize which edge state, and therefore which edge, has been excited.

\item \emph{Filtering.}\textemdash If there is only a single curve within a frequency range, then a mode exists for medium 1 over medium 2 but \emph{not} for its reverse. The non-simultaneous edge modes present in the $C_{3v}$ nontrivial case, Fig. \ref{fig:C3v_DC}, provide an example. These types of systems are utilized for filtering (see Sec. \ref{sec:filtering}).

\item \emph{Fourier space separation.}\textemdash This property has been alluded to countless times with regards to interfacial and cavity guide modes. When a source is placed at the start of a waveguide, the backscatter is inversely related to the wavelength of the energy-carrying envelope \cite{loncar_methods_2001}. Therefore, a larger wavevector is less prone to backscattering and hence less prone to coupling with its opposite propagating counterpart. 

\item \emph{Chirality of valley states and phase matching.}\textemdash Unique to topological valley modes is presence of a favoured chirality for edge modes \cite{xiao_valley-contrasting_2007-1, fefferman_edge_2015, qian_theory_2018}. The lack of coupling of modes with opposite chirality has been shown in \cite{xiaoxiao_direct_2017}. Therefore for a network, if a mode on a pre-nodal lead is of a particular chirality it will not easily couple to its counterpart of opposite chirality. This is also true for a mode with a particular phase, pre- and post- the nodal region. Hence, the chirality and phase matching properties are significant for determining the coupling between modes, pre- and post- the nodal region. 

\item \emph{Tunneling.}\textemdash A route to partitioning energy away from an interfacial waveguide is via the tunneling of energy through the decaying tails of the edge state. The 
amount of energy partitioned via tunneling is tuned by adjusting the band-gap.



\item \emph{Nodal region.}\textemdash The nodal region becomes highly relevant when the wavelength of the energy-carrying envelope is comparable in size to the nodal region. In these instances the design of the nodal region can preferentiate certain outgoing leads over others.
\end{itemize}

\subsection{Different modal shapes}
\label{sec:modal}

\begin{figure}[h!]
 \centering
 \begin{tabular}{ll}
	\includegraphics[scale=0.105]{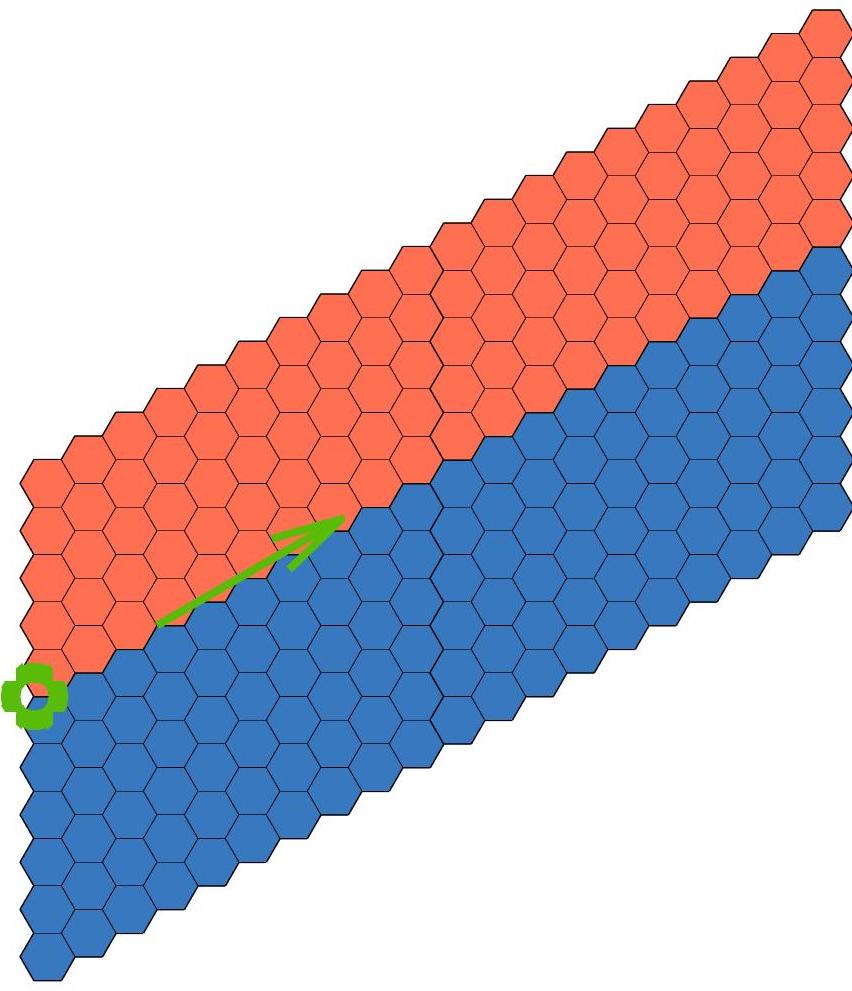}
 		\includegraphics[scale=0.15]{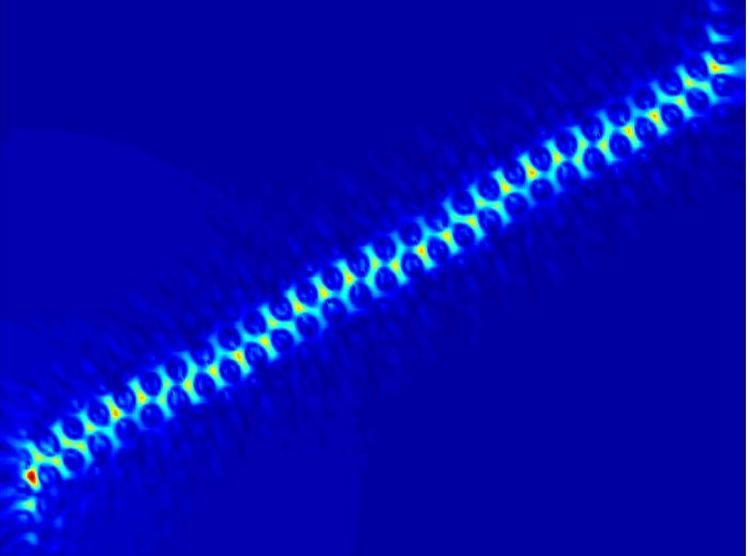}
\end{tabular}
\caption{The clarity of the ``sawtooth" edge mode obtained, for parameter values from the convex curve ($\omega=15.67$) shown in Fig. \ref{fig:C6v_DC}, is evident for this $C_{6v}$ nontrivial example.}
\label{fig:C6v_ZLM_OB}

 \end{figure}

\begin{figure}[h!]
 \centering
 \begin{tabular}{ll}
\includegraphics[scale=0.115]{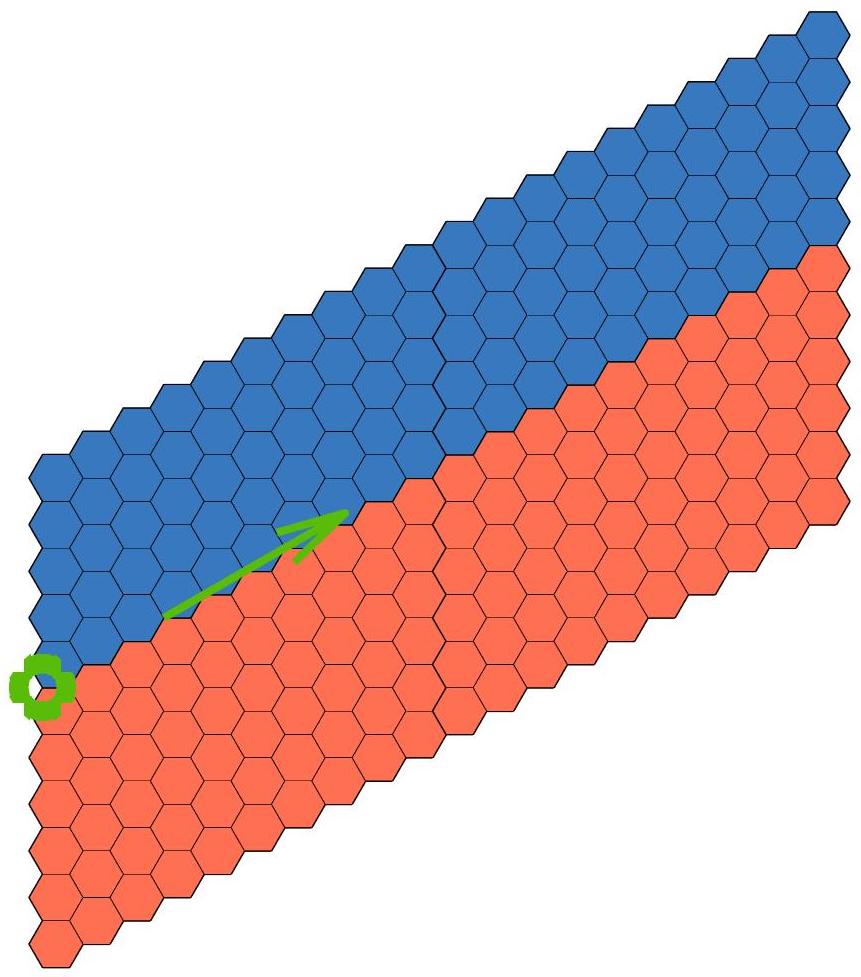}
 		\includegraphics[scale=0.1675]{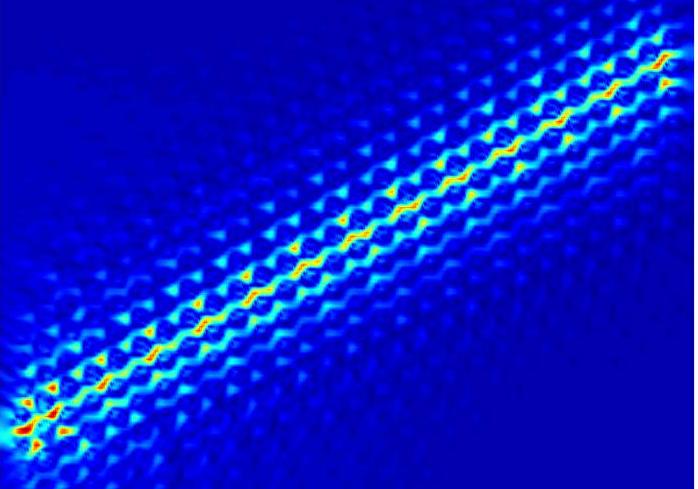}	
\end{tabular}
\caption{Edge state for the same parameters as Fig. \ref{fig:C6v_ZLM_OB}, but now with the ordering of the media reversed to be associated with the concave curve in Fig. \ref{fig:C6v_DC}. The modal pattern is clearly different from the ZLM in Fig. \ref{fig:C6v_ZLM_OB}.}
\label{fig:C6v_ZLM_BO}
 \end{figure}

The clarity of edge modes that we find numerically allows us to easily, and rapidly, identify whether the edge state corresponding to, say, the concave or convex curves of Fig. \ref{fig:C6v_DC} is excited. 
 The two edge modes shown in Figs \ref{fig:C6v_DC}, \ref{fig:C3v_DC} and \ref{fig:C3v_nonorthog_DC} relate to either an edge state along medium 1 over medium 2 or its reverse. They are visualised by placing a source at the far-left hand side of the interface between the two media; the resulting modal pattern clearly reflects the relative ordering of the media, see Figs \ref{fig:C6v_ZLM_OB} and \ref{fig:C6v_ZLM_BO}. 


For the simpler cases,  Figs \ref{fig:C6v_ZLM_OB}, \ref{fig:C6v_ZLM_BO}, the identity of the edges is obvious, however later
using the highly resolved edge modes to unpick which edge is responsible, and exactly which mode is excited, is in practice very useful when constructing complex networks comprised of many geometrically distinct regions. The underlying mathematics that underpins the construction of interfacial wave networks does not rely upon the physical model and so there is no need to use a more complicated systems, i.e. Maxwell, acoustic, Navier elasticity, than the K-L flexural plate equation.
 These effects are geometrically induced, hence system-independent, choosing more complicated models to explore them adds nothing more than computation time and results in lower resolution edge states which obscure from the fundamental physics. By using the Kirchhoff-Love model, as a vehicle, we are able to use the resulting clarity of the modes to provide us with useful information in a time-efficient manner.

\begin{figure}[h!]
 \centering
 \begin{tabular}{ll}
 \includegraphics[scale=0.205]{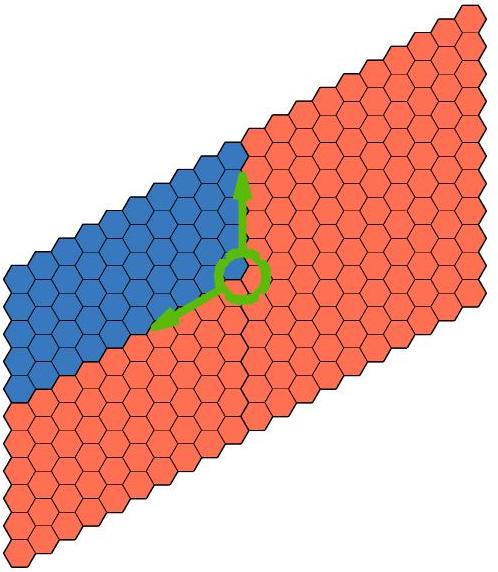}
 \includegraphics[scale=0.1305]{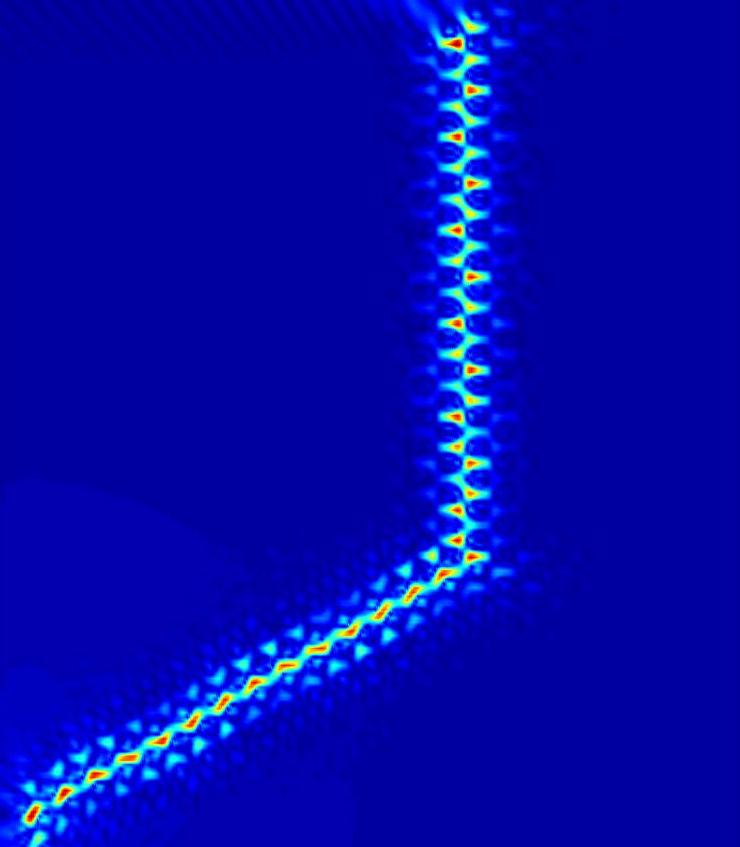}
\end{tabular}
\caption{Source placed at the turning point, $\omega = 15.80$. As evident from the modal shapes, the mode propagating upwards is the ``sawtooth" mode that lies on the convex curve whilst the leftward propagating mode is on the concave curve as shown in Fig. \ref{fig:C6v_DC}}
\label{fig:C6v_bend}
 \end{figure}

A less trivial example than Figs \ref{fig:C6v_ZLM_OB}, \ref{fig:C6v_ZLM_BO}, touched upon in refs. \cite{makwana_geometrically_2018, liu_tunable_2018}, is that of a gentle waveguide bend where the adjoining media undergoes a $2\pi/3$ bend, see Fig \ref{fig:C6v_bend}. A source placed at the turning point between the two interfaces excites either the mode belonging to the concave or convex curves;  the clarity of the modes shows clearly the origin of the leftward and upward propagating modes. 

\subsection{Filtering}
\label{sec:filtering}

\begin{figure}[h!]
 \centering
 \begin{tabular}{ll}
	\includegraphics[scale=0.115]{ZLM_BO.jpg}
	\includegraphics[scale=0.165]{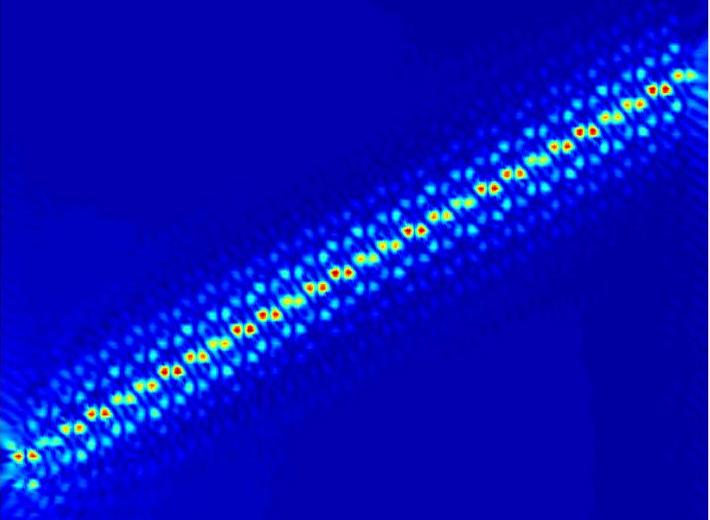}
\end{tabular} 
\begin{tabular}{ll}
	\includegraphics[scale=0.115]{ZLM_OB.jpg}
	\includegraphics[scale=0.165]{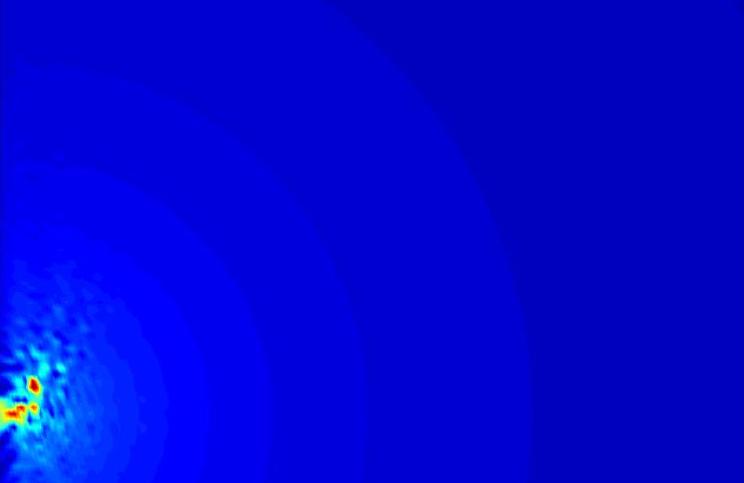}
\end{tabular}
\caption{The importance of the relative ordering of the media for filtering in the $C_{3v}$ nontrivial case, for both panels an isotropic source is placed at the leftmost edge at frequency $\omega = 19.13$.} 
\label{fig:filtering_ex}
 \end{figure}

The $C_{6v}$ case, referenced in the previous section, has two distinct broadband zero-line modes (ZLMs), at overlapping frequencies, 
within the nontrivial band gap (Fig. \ref{fig:C6v_DC}); the asymmetry of the edges was reflected by the differences in the modal shapes. In contrast, for the $C_{3v}$ case, Fig. \ref{fig:C3v_DC}, the ZLMs now have very limited overlap with only one broadband mode. Physically, this implies that, for the $C_{3v}$ case, a ZLM exists, over a wide range of frequencies, for one of the orderings of the media but not for its inverse. This allows us to restrict propagation along one of two distinct interfaces; see Fig. 
 \ref{fig:filtering_ex}.

\subsection{Fourier space separation}
\label{sec:Fourier_space}

\begin{figure}[h!]
 \centering
 \begin{tabular}{ll}
	\includegraphics[scale=0.17]{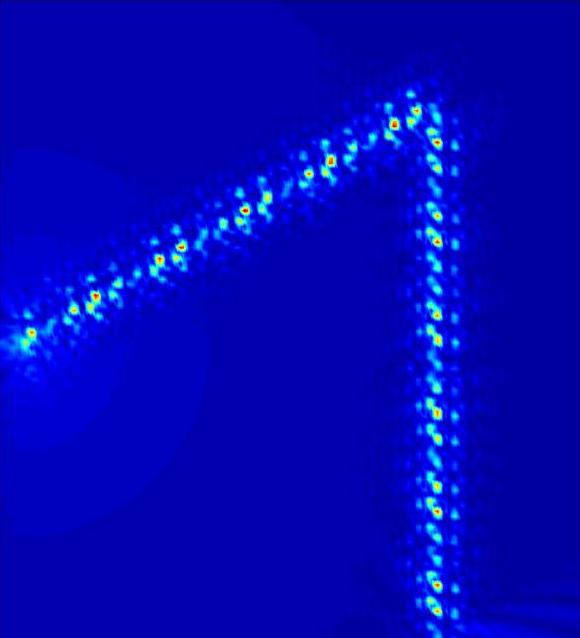}
	\includegraphics[scale=0.17]{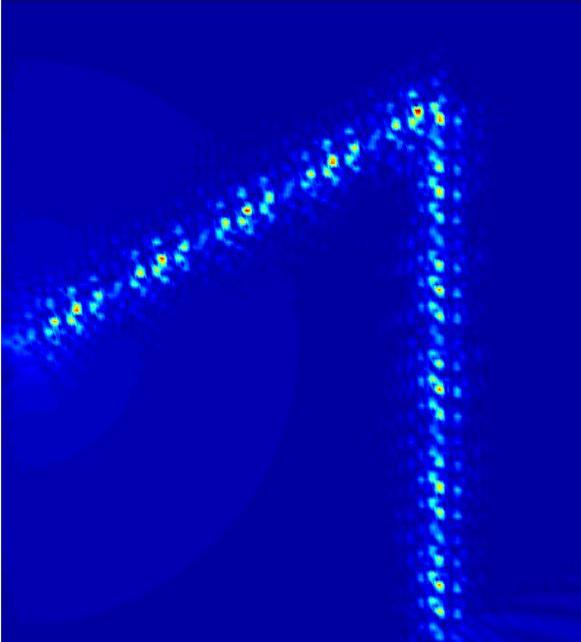}
\end{tabular} 
\begin{tabular}{ll}
	\includegraphics[scale=0.18]{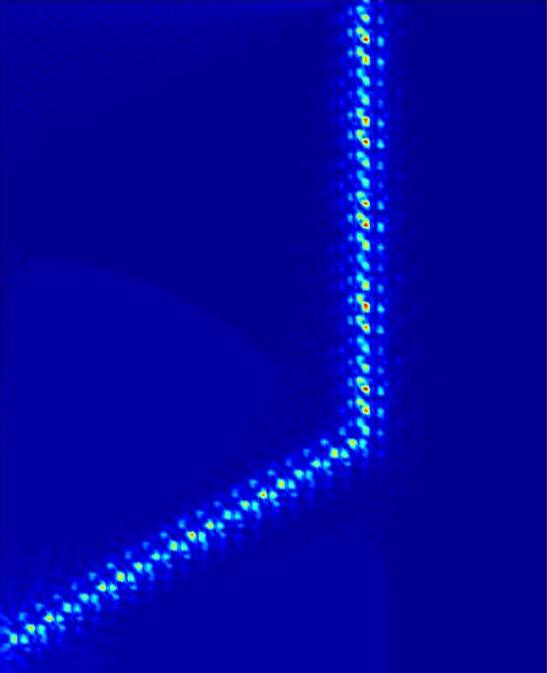}
	\includegraphics[scale=0.18]{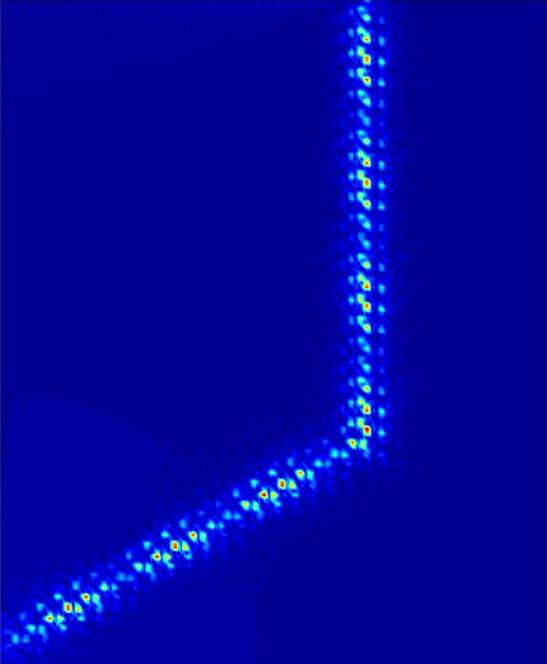}
\end{tabular}
\caption{Robustness of $C_{3v}$ trivial edge state demonstrated against $\pi/3$ and $2\pi/3$ bend. The source excitation is placed at the leftmost edge.}
\label{fig:trivial_sharp_bend}
 \end{figure}

It is perhaps surprising, given the emphasis in the topological literature on nontrivial edge states, to observe that the trivial case demonstrates visually robust energy transport around $\pi/3$ and $2\pi/3$ bends (see Fig.  \ref{fig:trivial_sharp_bend}). This is because  the transport around the bends is partly supported by the separation in Fourier space between the forward and backward propagating modes; this is also implicit in the successful guiding of photonic crystal waveguides around bends  \cite{loncar_methods_2001} where topological protection is also absent. 
 An example showing the importance of the separation, common in the valleytronics literature, is 
the contrast between the armchair and zigzag interfaces. This is further evidenced by the small Fourier separation between modes of opposite group velocity, see \cite{bi_role_2015}, for the armchair case relative to that of the zigzag; the armchair termination is far more prone to backscatter than the zigzag.

\subsection{Chirality of valley states and phase matching}
\label{sec:chirality_phase}
\begin{figure}[h!]
 \centering
  \begin{tabular}{ll}
	(a)	$C_{3v}$ trivial, $v_g < 0$ & (b) $C_{3v}$ trivial, $v_g > 0$
\end{tabular}
 \begin{tabular}{ll}
	\includegraphics[scale=0.16]{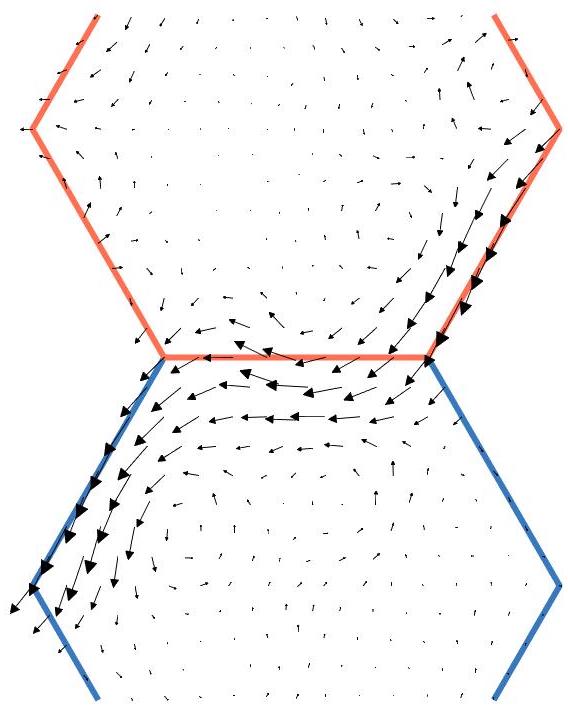}
\includegraphics[scale=0.16]{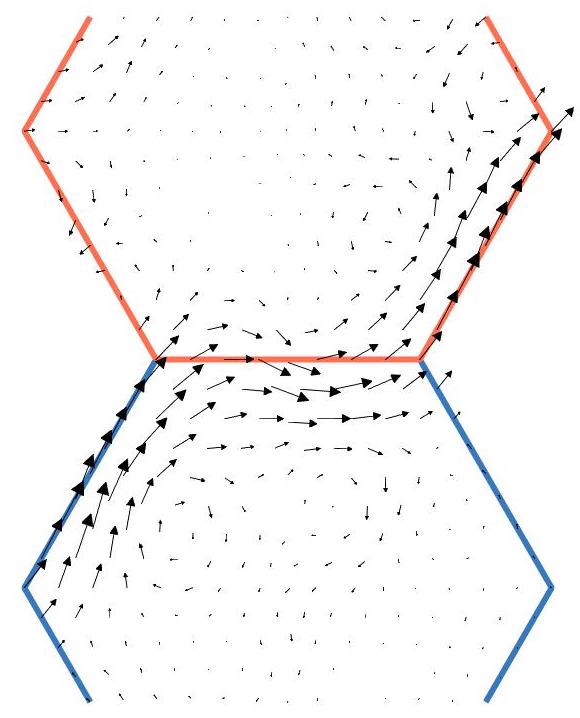}	
\end{tabular}
\begin{tabular}{ll}
(c) $C_{6v}$ nontrivial, $v_g < 0$ & (d) $C_{6v}$ nontrivial, $v_g > 0$
\end{tabular}
\begin{tabular}{ll}
		\includegraphics[scale=0.16]{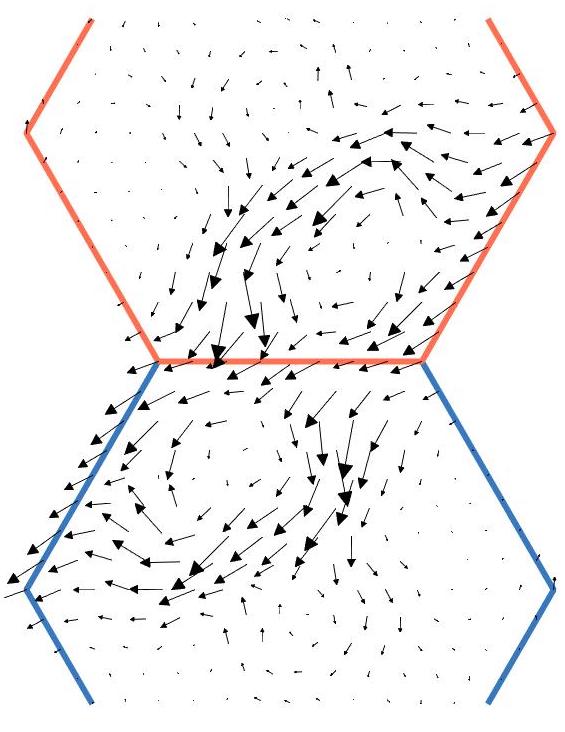}
		\includegraphics[scale=0.16]{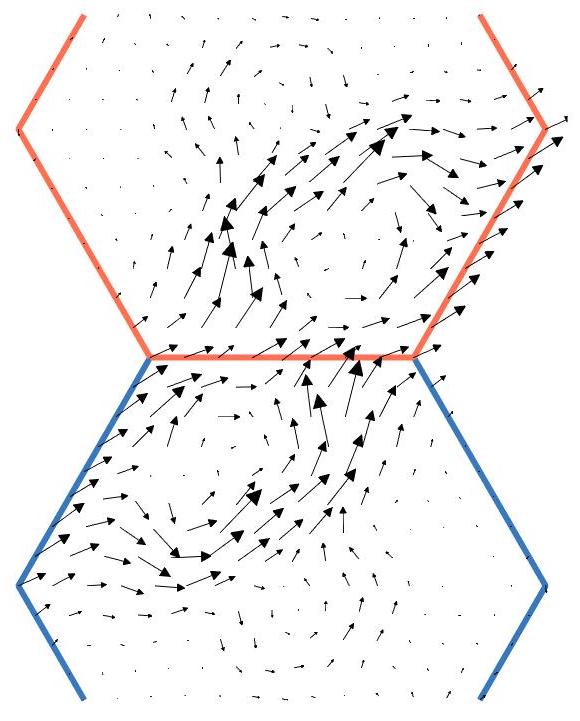}
\end{tabular}
\caption{Close-ups of  the interfaces for the $C_{3v}$ trivial and $C_{6v}$ nontrivial cases; top and bottom panels, respectively. 
The left and right panels represent negative and positive group velocity, $v_g$, respectively.
The arrows denote the energy flux, Eq. \eqref{eq:flux}. There is a clear distinction between the topologically nontrivial and trivial cases; the nontrivial flux has orbital motion induced from the stacked media having opposing valley Chern numbers and hence opposite chirality at the $KK'$ valleys; this property that gives the valley modes their robustness \cite{fefferman_edge_2015, fefferman_bifurcations_2016}.}
\label{fig:chirality}
 \end{figure}

To ensure coupling between modes, pre- and post- the nodal region, we must consider the relative group velocity and phases of the incoming and outgoing waves. The time-averaged energy flux, for a structured elastic plate \cite{norris_scattering_1995}, 
\beq
\langle {\bf F} \rangle = \frac{\omega_{\bkappa}}{2} \text{Im} \left( \psi_{j\bkappa} \nabla_{\bx}^3 \psi^*_{j\bkappa} - \nabla_{\bx}^2 \psi^*_{j\bkappa} \nabla_{\bx} \psi_{j\bkappa} \right), 
\label{eq:flux}
\eeq
 provides the natural quantity that describes the energy transfer.  
The topological protection of the edge states arises from the orbital nature of their flux; there is a clear difference in the fluxes between the topologically nontrivial and trivial edge states as evidenced by Fig. \ref{fig:chirality}. The relative difference in their robustness is explained more rigorously in \cite{fefferman_edge_2015, fefferman_bifurcations_2016}

An additional impact of the orbital flux is that modes of opposite chirality do not couple \cite{cheng_robust_2016-1}; hence topological networks have to be carefully designed in order to trigger the desired excitations along outgoing leads. Another crucial condition that dictates the coupling between pre- and post-nodal region modes is the phase. Transmission across a nodal region will be facilitated when the phase of the incoming and outgoing waves match; this is commonly known as phase matching. 

Both the phase and energy flux of the post-nodal region modes must match the pre-nodal mode in order for the wave to propagate through with limited scattering. Topologically nontrivial systems that require matching of the phase and chirality, between incoming and outgoing leads, are often said to suppress intervalley scattering \cite{chen_defect_2009, morpurgo_intervalley_2006, pesin_spintronics_2012, morozov_strong_2006}. For the $C_{6v}$ nontrivial case (and every other topological case \cite{cheng_acoustic_2016, xiaoxiao_direct_2017, zhang_manipulation_2018}) this restricts the systems to a 2-way splitting of energy (if a tunneling mechanism is not invoked, see Sec. \ref{sec:tunneling}).

\begin{figure} [h!]
	\includegraphics[scale=0.195]{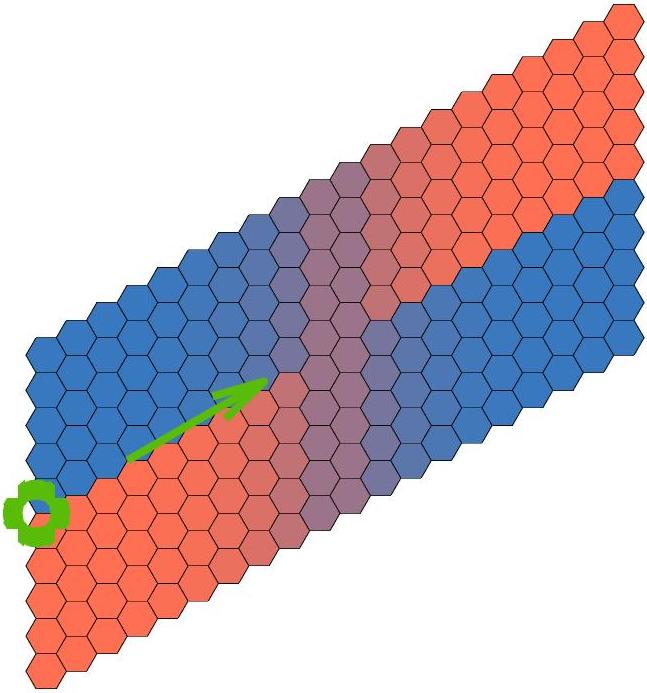}
\caption{Schematic for the graded medium associated with examples shown in Fig. \ref{fig:graded_nontrivial} and \ref{fig:graded_trivial}. Source placed at leftmost edge, we grade both, $C_{6v}$ nontrivial and $C_{3v}$ trivial examples.
The former is graded according to the relative difference in value between alternate masses (see top panel of Fig. \ref{fig:graded_nontrivial}) whilst the latter is graded according to the angular perturbation away from $\sigma_v$ (see top panel of Fig. \ref{fig:graded_trivial}). Propagating modes lie within the graded region in the center.}
\label{fig:graded_schematic}
\end{figure}

\begin{figure} [h!]
\centering
\caption{Graded $C_{6v}$ nontrivial case}

	\includegraphics[scale=0.1325]{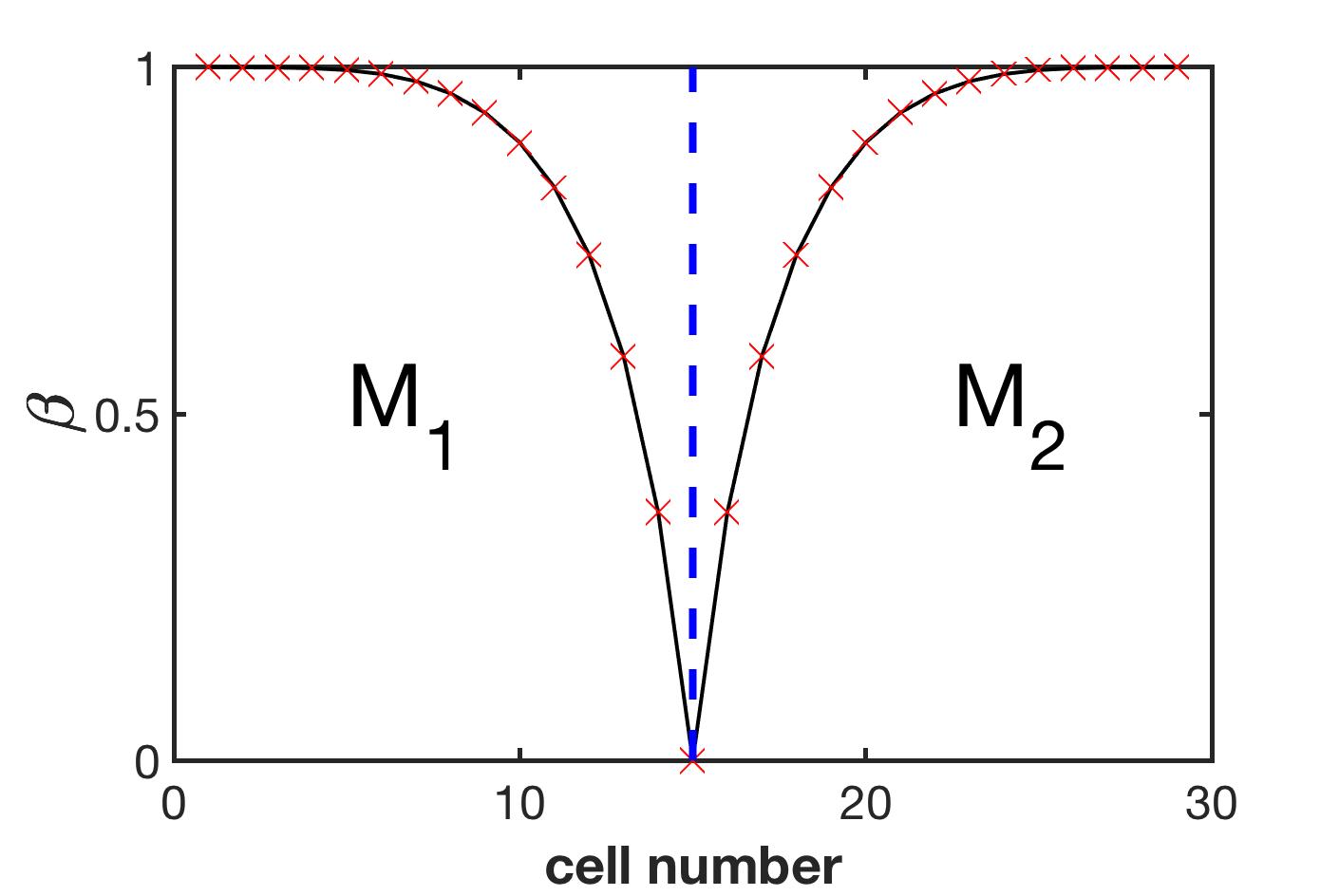}
	
	Graded variation of mass for upper medium in Fig. \ref{fig:graded_schematic}. Left of the vertical dashed line, denotes the $\beta$ perturbation away from $M_0$ for the mass $M_1$, whilst right of the dashed line represents the $\beta$ perturbation away from $M_0$ for $M_2$. The red crosses represent the discrete mass values taken within a cell. Lower medium variation is identical albeit from right to left.
	\label{fig:graded_mass}
	\includegraphics[scale=0.165]{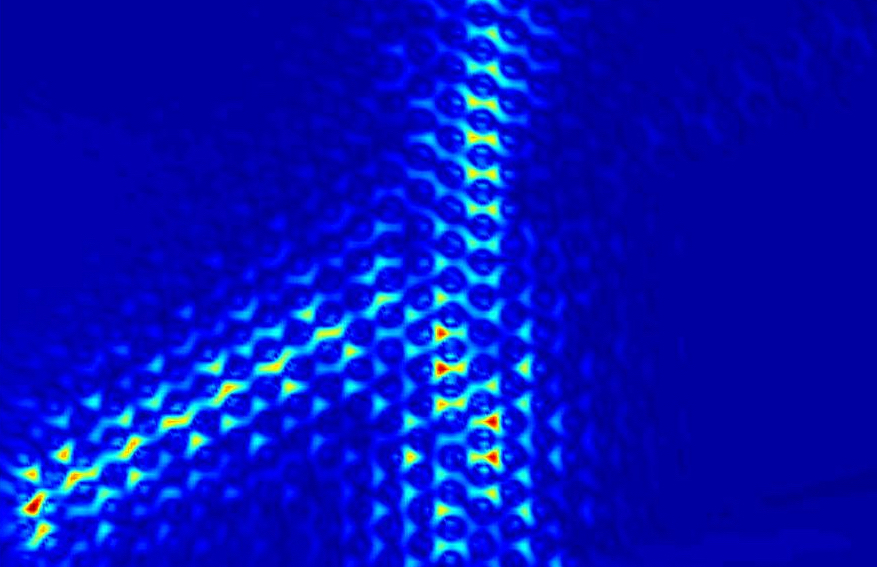}
	
	Scattered field demonstrates lack of coupling between the pre-graded region ZLM to the post-graded region mode. This is due to the matching chirality mode being located at the opposite valley in Fourier space (Fig. \ref{fig:C6v_DC}).
%
\label{fig:graded_nontrivial}
\end{figure}

\begin{figure} [h!]
\centering
\caption{Graded $C_{3v}$ trivial case}
	\includegraphics[scale=0.1325]{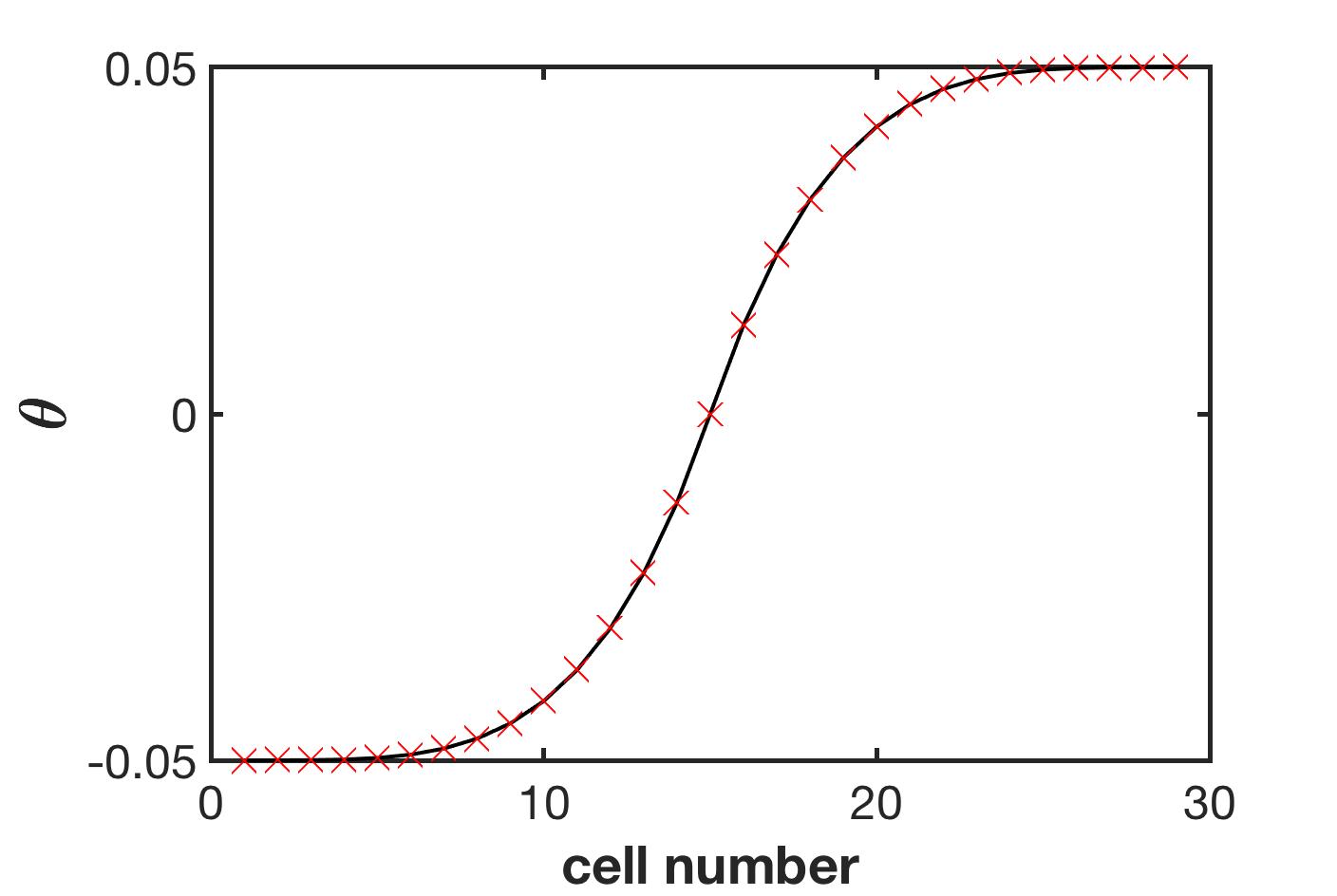}
	
	Graded variation in angular perturbation for upper medium in Fig. \ref{fig:graded_schematic}. $\theta$ represents the perturbation away from the vertical reflectional symmetry line, $\sigma_v$ (Fig. \ref{fig:edges}), for the $C_{3v}$ case. The red crosses denote the discrete values taken within a cell. Lower medium variation is identical albeit from right to left.
	\label{fig:graded_angle}
	\includegraphics[scale=0.285]{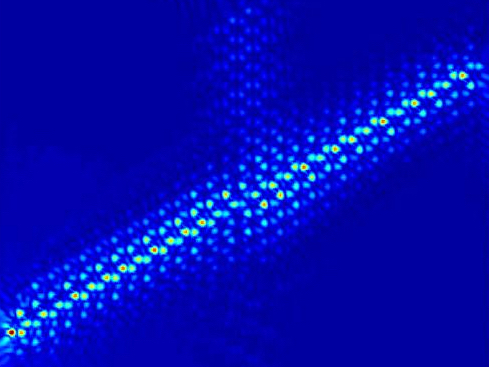}
	
	Scattered field demonstrates coupling between the pre-graded and post-graded region modes. This is due to the matching phase and group velocity of the modes shown in Fig. \ref{fig:C3v_nonorthog_DC}.
\label{fig:graded_trivial}
\end{figure}

A continuous spatial change of medium so the geometry changes with distance can be used to illustrate many of the features that are important, for instance the  absence of coupling, for the $C_{6v}$ nontrivial case, and coupling, for the $C_{3v}$ trivial case, shown in Figs. \ref{fig:graded_schematic}, \ref{fig:graded_nontrivial}, \ref{fig:graded_trivial}. Fig. \ref{fig:graded_schematic} pictorially shows the graded change from one media ordering to its reverse. The precise grading function for the upper medium (Fig. \ref{fig:graded_schematic}) in the $C_{6v}$ case is shown in the upper panel of Fig. \ref{fig:graded_nontrivial}; the lower medium variation is identical albeit from right to left. The use of material grading allows us to explore and demonstrate the coupling mechanism, the resulting scattered field is shown in the lower panel of Fig. \ref{fig:graded_nontrivial}. Evidently the left hand ZLM is unable to couple into the right hand ZLM; this is due to the post-graded region mode being located at the opposite valley, $-\bkappa$, to the pre-graded region mode $+\bkappa$ (see dispersion curves in Fig. \ref{fig:C6v_DC}), hence the absence of phase matching impedes the propagation through the graded region. In contrast, the $C_{3v}$ trivial case exhibits modes, of identical group velocity and phase, for the pre- and post-graded region leads (see dispersion curves in Fig. \ref{fig:C3v_nonorthog_DC}); this results in almost unimpeded propagation through the graded region, Fig. \ref{fig:graded_trivial}. The grading, for the upper medium in the $C_{3v}$ trivial case, is shown in the upper panel of Fig. \ref{fig:graded_trivial}; similar to the $C_{6v}$ nontrivial case, the lower mediums' grading is the reverse of the upper medium. The primary visual differential between the curves in Figs. \ref{fig:C6v_DC} and \ref{fig:C3v_nonorthog_DC}, that results in this different propagative behaviour, is the curvature of the pair of edge states; the two convex curves in Fig. \ref{fig:C3v_nonorthog_DC} ensure the matching of group velocities at a particular $\bkappa$ whilst the concave and convex curves in Fig. \ref{fig:C3v_nonorthog_DC} do not.


Despite the $C_{3v}$ case being trivial, its relative robustness against sharp disorders (demonstrated by Fig. \ref{fig:trivial_sharp_bend}) and its additional coupling capabilities, relative to its nontrivial counterpart, allow for the construction of novel interfacial wave networks that differ from the topologically nontrivial examples \cite{cheng_acoustic_2016, xiaoxiao_direct_2017, zhang_manipulation_2018}; specifically, the ability to have more than 2-way energy splitters (Sec. \ref{sec:trivial_networks}). However, the negative of using these trivial modes, for networks, is the prevalence of scattering and hence lack of controllability compared with the nontrivial cases; for the latter, the conservation of topological charge post- the nodal region \cite{ezawa_topological_2013} leads to more robust edge states along the outgoing leads and greater tunability in partitioning energy. In order to construct topological networks, that contain more than 2-way energy-splitters, the tunneling mechanism must be used as in Secs. \ref{sec:tunneling}, \ref{sec:tunnel_network}.

\subsection{Tunneling}
\label{sec:tunneling}
An alternative means to transmit energy along different leads is via tunneling; the exponentially decaying tail, of an incoming mode, being used to ignite an outgoing mode. This allows for transmission of energy down leads that would not be activated due to a mismatch in phase and/or chirality (see Sec. \ref{sec:chirality_phase}). 
 The benefit of utilizing tunneling to redirect energy away from an incoming ZLM is that this enables more than 2-way energy-splitting (see Sec. \ref{sec:tunnel_network}) whilst still preserving the topological charge, and hence the topological protection, along the post-nodal region leads. 

 The width of the band-gap, as mentioned in \cite{qian_theory_2018}, plays a role: The band-gap needs to be small enough to preserve a strong Berry curvature in the vicinity of the valleys, but large enough to enhance the localization of modes along the interface. The former criterion, is related to the chirality of the edge modes, see Sec. \ref{sec:chirality_phase}, whilst the latter is related to the decay perpendicular to the propagation direction. If these criteria are balanced then the topological protection of nontrivial states and the localization of the states is optimized. The latter is important for tuning the partitioning of energy away from the nodal region.

\begin{figure} [htb!]
 \centering
 \begin{tabular}{ll}
	\includegraphics[scale=0.135]{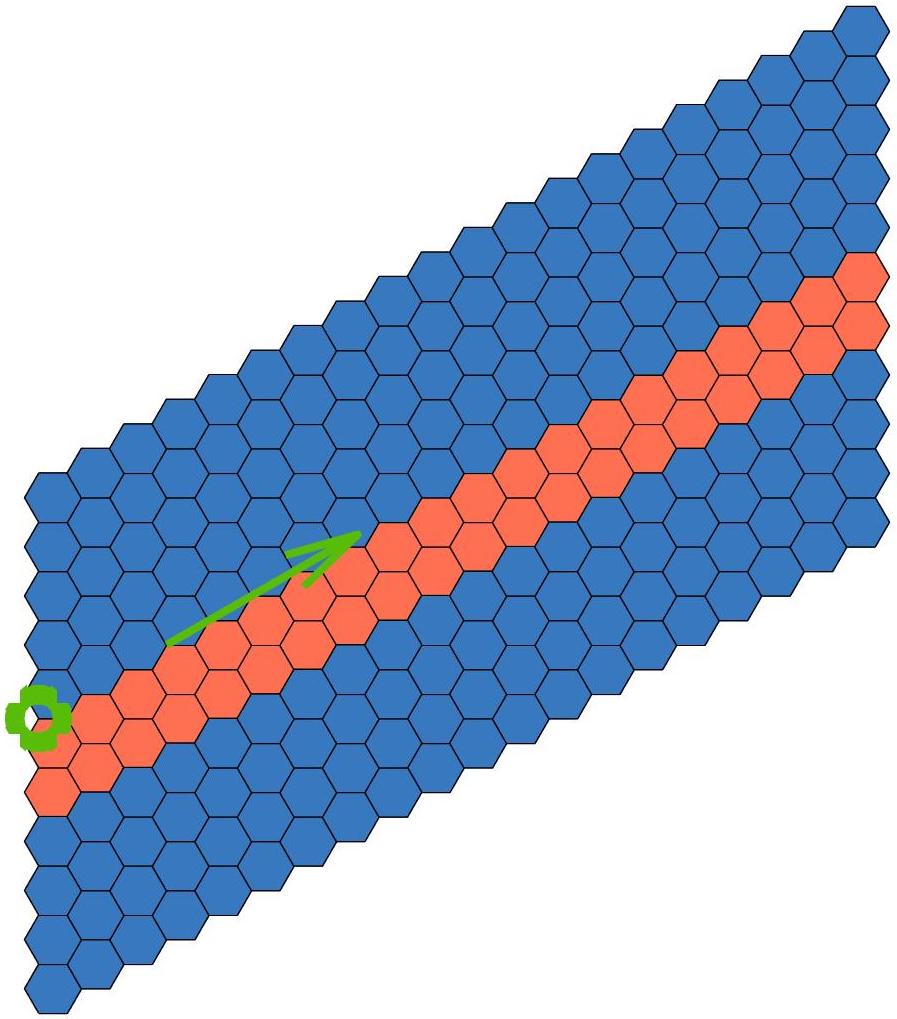}
\label{fig:tunnel_schem} 
		\vspace{0.5cm}
	\includegraphics[scale=0.135]{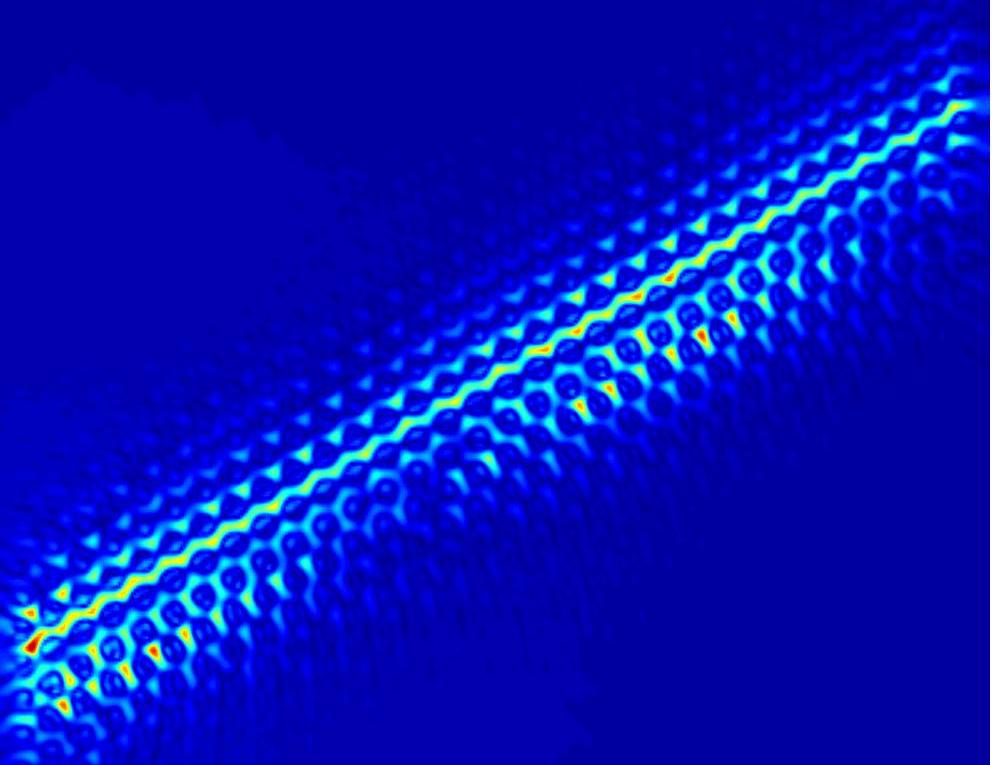}
\label{fig:tunnel_ex}
\end{tabular}
\caption{Demonstration of tunneling for the $C_{6v}$ nontrivial case. A source is placed along the upper interface where $\omega = 15.375$; the decaying tail of the triggered ZLM ignites the parallel ZLM.}
\label{fig:tunneling_example}
\end{figure}

A simple example of tunneling, for the $C_{6v}$ case is shown in Fig. \ref{fig:tunneling_example}; the ignited ZLM has the modal pattern shown in Fig. \ref{fig:C6v_ZLM_BO}, this mode tunnels to ignite the parallel ZLM which has the sawtooth pattern shown in Fig. \ref{fig:C6v_ZLM_OB}.
The back and forth coupling between two evanescently coupled  parallel modes is a well-known phenomenon for parallel waveguides classically treated using coupled mode theory \cite{yariv_coupled-mode_1973} and its variants \cite{hammer_hybrid_2007}.  

\subsection{Nodal region}
\label{sec:nodal}

\begin{figure} [h!]
\centering
\begin{tabular}{ll} 
	\includegraphics[scale=0.170]{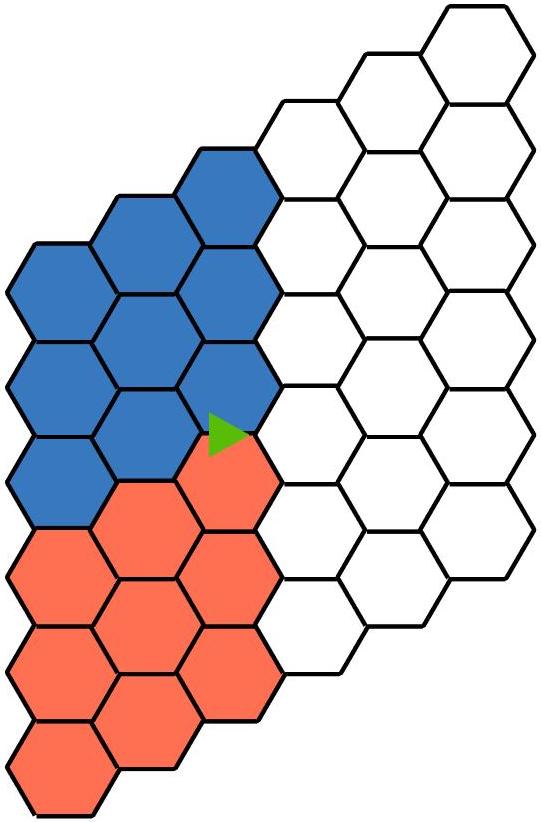}
	\label{fig:nodal_region_overview}
	\includegraphics[scale=0.125]{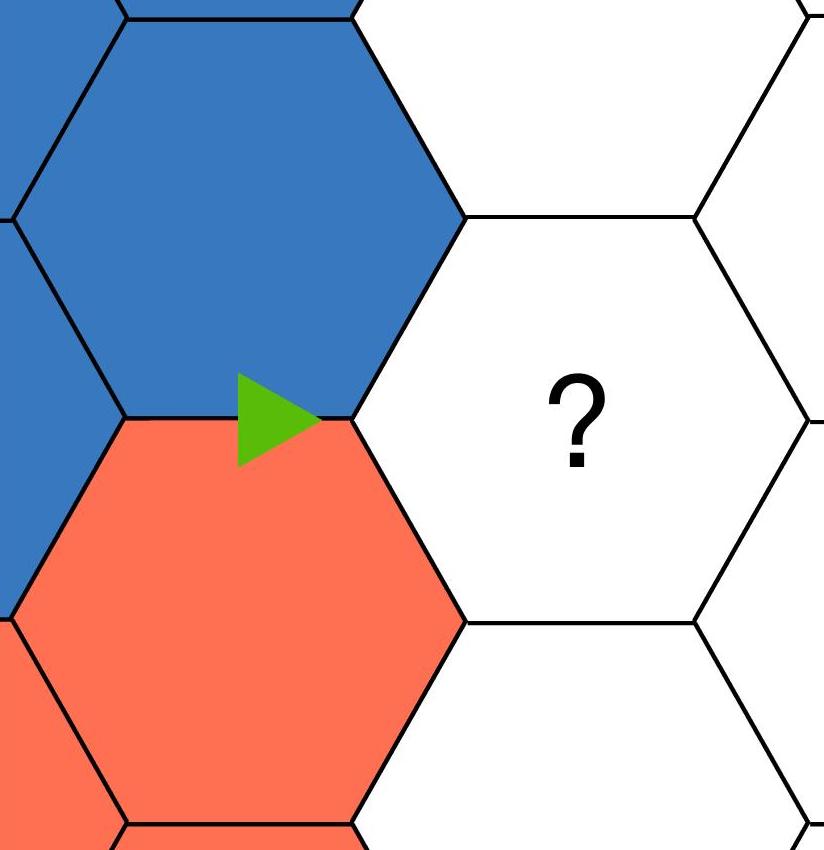}
		\label{fig:nodal_region_zoomin}
\end{tabular}
\caption{Incoming interfacial mode from the left lead, two right-sided outgoing leads. There are many ways to partition the right-sided medium  into an even number of regions. If wavelength is comparable to size of the nodal region the partitioning of energy will be very sensitive to the design of the nodal region.}
\label{fig:nodal_region}
\end{figure}

The design of the nodal region impacts upon the relative transmission along active leads, where propagation is permitted, and hence is an important property when designing interfacial wave networks particularly when wavelength and nodal region are comparable in scale. These networks are constructed upon a medium which contains an even number of geometrically distinct portions; an interfacial mode is able to propagate between each pair of distinct media. If we are dealing with a nodal region that divides four media, for example, there are a myriad of ways to design it; each of which will partition energy differently especially when the wavelength of the incoming wave is comparable to the nodal region. Fig. \ref{fig:nodal_region} illustrates this clearly; the hexagonal cell has one incoming edge and two outgoing edges (right panel of Fig. \ref{fig:nodal_region}) and yet we wish to partition energy in 2 or 3 directions using non-fractional cell partitions between the distinct media. For topological networks, the chirality and phase matching arguments, readily used, determine which leads energy can travel down but the relative transmission down those active leads remains unknown. This is seen in  the topological network designs of \cite{cheng_acoustic_2016, xiaoxiao_direct_2017} where there is a difference in transmission between the active leads; additional information is garnered by considering the nodal region. These particular examples are examined more closely in Sec. \ref{sec:trivial_networks}. 

\section{Building Novel Topological Networks}
\label{sec:building} 

The knowledge accrued, in the previous section, regarding the transport of energy is essential for building complex interfacial wave networks. In this section, we geometrically engineer networks that have additional functionality as compared to the current designs in the valleytronics literature. We employ, both, the topologically nontrivial and trivial examples to yield designs which go beyond 2-way energy-splitting (Secs. \ref{sec:trivial_networks}, \ref{sec:tunnel_network}). Section \ref{sec:supernetwork}, uses the building blocks of the design paradigm, described in the previous section, to produce the first realization of a topological supernetwork. 

\subsection{Filtering: restricting propagation using $C_{3v}$ nontrivial modes}
\label{sec:filter_network}

\begin{figure} [htb!]
\centering
\begin{tabular}{llllll} 
(a)	\includegraphics[scale=0.090]{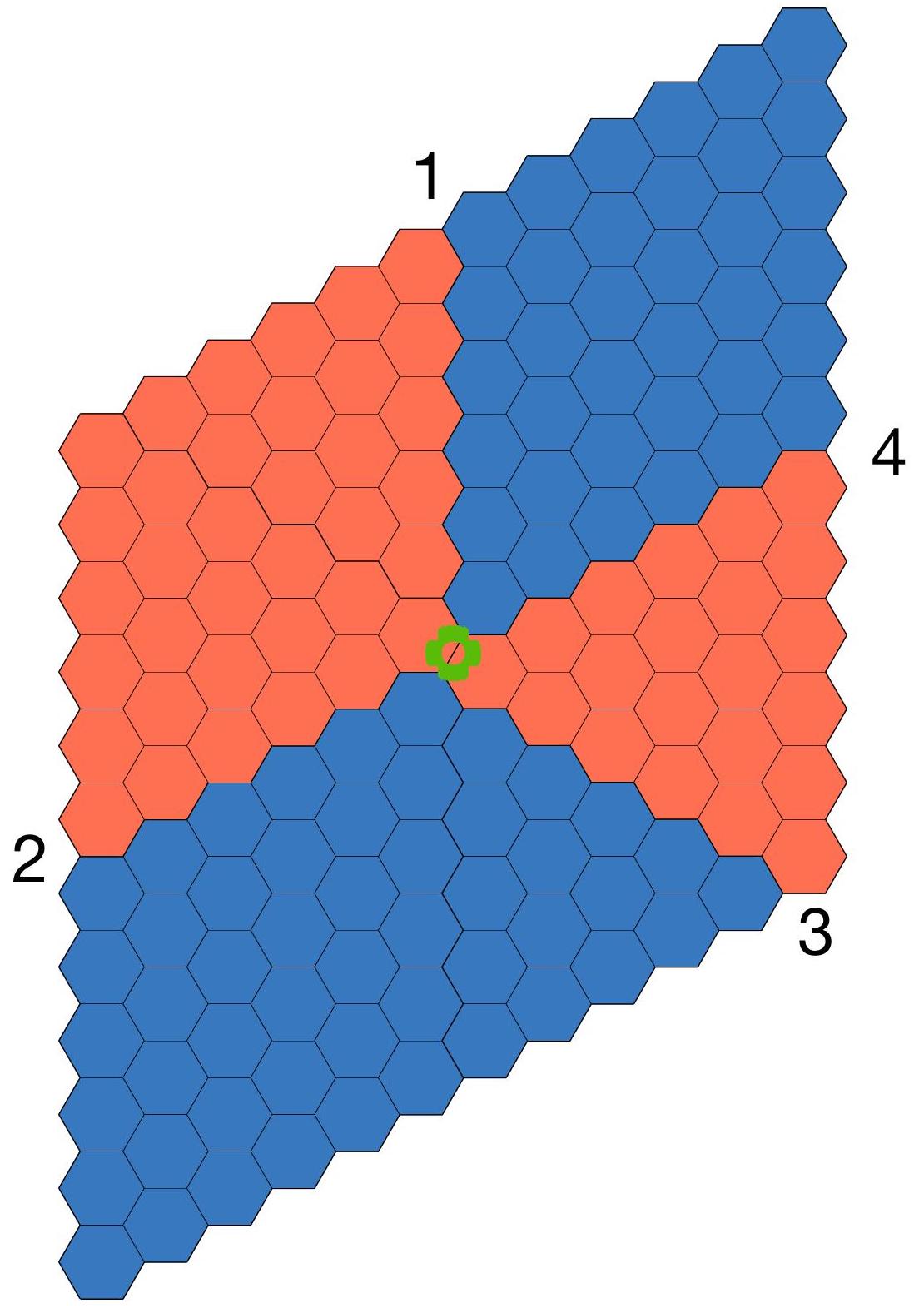}
	\label{fig:arrow_schematic}
(b)	\includegraphics[scale=0.335]{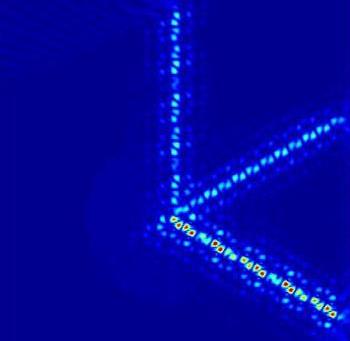}
\label{fig:C3v_arrow}
(c) 
	\includegraphics[scale=0.300]{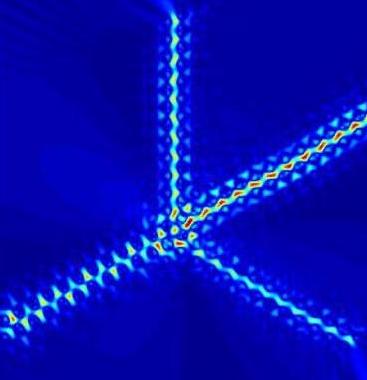}
\label{fig:C6v_arrow}
\end{tabular} 
\caption{Filtering network: (a) schematic, (b) $C_{3v}$ nontrivial case where the edge state along interface 2 is not excited, $\omega = 18.93$, and (c) $C_{6v}$ where the ``sawtooth" mode along interface 2 is clearly triggered, $\omega = 15.91$. $1152$ hexagonal cells are used for all the networks in this section.}
\label{fig:filter_network}
\end{figure}

The $C_{3v}$ nontrivial case demonstrates the filtering properties, described earlier in Sec. \ref{sec:filtering}, in Fig. \ref{fig:filter_network}. We place a dipole between the two central blue cells, Fig. \ref{fig:filter_network}(a); each source has the potential to trigger 
all of the ZLMs $1-4$; for the $C_{3v}$ nontrivial case, despite ZLMs $1,3$ and $4$ being triggered, ZLM $2$ is not directly excited, Fig. \ref{fig:filter_network}(b). This is due to the interface of ZLM $2$, closest to the dipole, being associated to the narrowband mode and hence different from the other three. The difference in interfaces is revealed by replacing the $C_{3v}$ topologically nontrivial domains with those from the $C_{6v}$ nontrivial case; from Fig. \ref{fig:filter_network}(c), the modal shape of ZLM $2$ is clearly distinct from ZLMs $1, 3$ and $4$ (Sec. \ref{sec:modal}); this difference visually validates the absence of ZLM $2$ for the $C_{3v}$ nontrivial case. The potential for geometrically induced filtering enhances our design capability through  tunability of energy propagation that can be restricted to only propagate along selected interfaces.

\subsection{Using $C_{3v}$ trivial modes for 3- and 5-way splitting }
\label{sec:trivial_networks}

\begin{figure} [h!]
\centering
\begin{tabular}{lll} 
(a) Schematic  \qquad    & \qquad (b) Topologically nontrivial 2-way energy-splitter \qquad & (c) Topologically trivial 3-way energy-splitter
\end{tabular}

\begin{tabular}{lll}
	\includegraphics[scale=0.215]{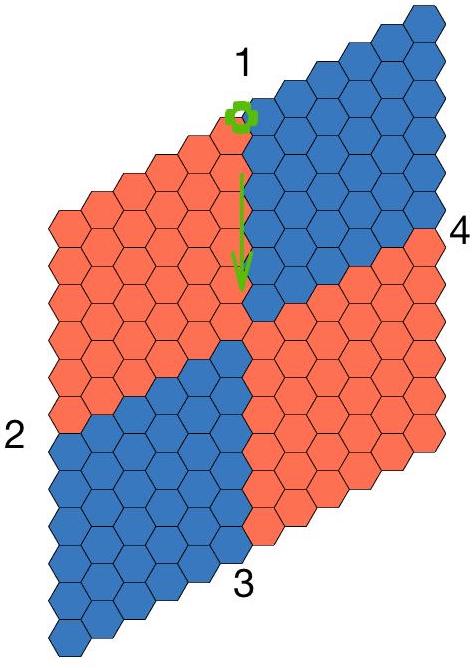} \hspace{1.0cm} &
	\label{fig:X}
	\includegraphics[scale=0.200]{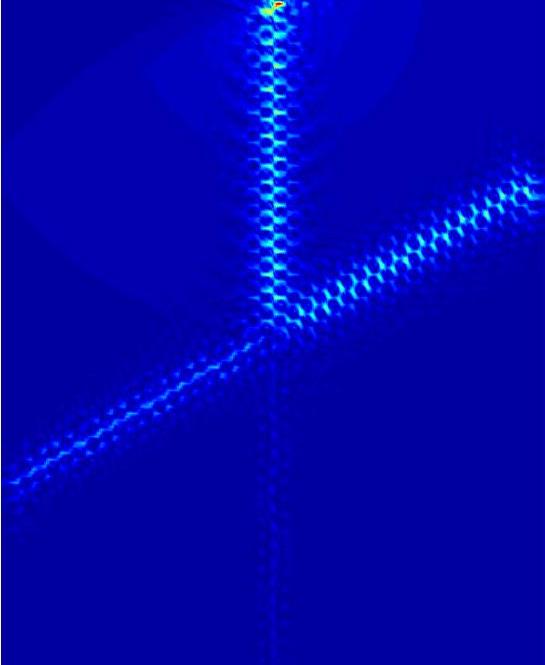} \hspace{0.5cm} &
\label{fig:2_splitter_X}
	\includegraphics[scale=0.285]{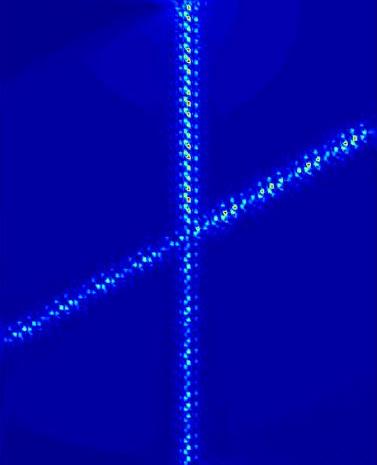}
\label{fig:3_splitter_X}
\end{tabular} 
\caption{Tesselation used for 3-way energy-splitting motivated by \cite{xiaoxiao_direct_2017}, panel (a); source is placed at the beginning of interface 1 and the resulting scattered fields for the $C_{6v}$ nontrivial and $C_{3v}$ trivial cases are shown in panels (b) and (c) respectively. Panel (b) resembles the displacement in \cite{xiaoxiao_direct_2017}, frequency $\omega = 15.86$; whilst panel (b) shows novel 3-way splitting of energy at $\omega = 19.45$.}
\label{fig:NatComm_Ex}
\end{figure}

\begin{figure} [h!]
\begin{tabular}{ll} 
	\includegraphics[scale=0.150]{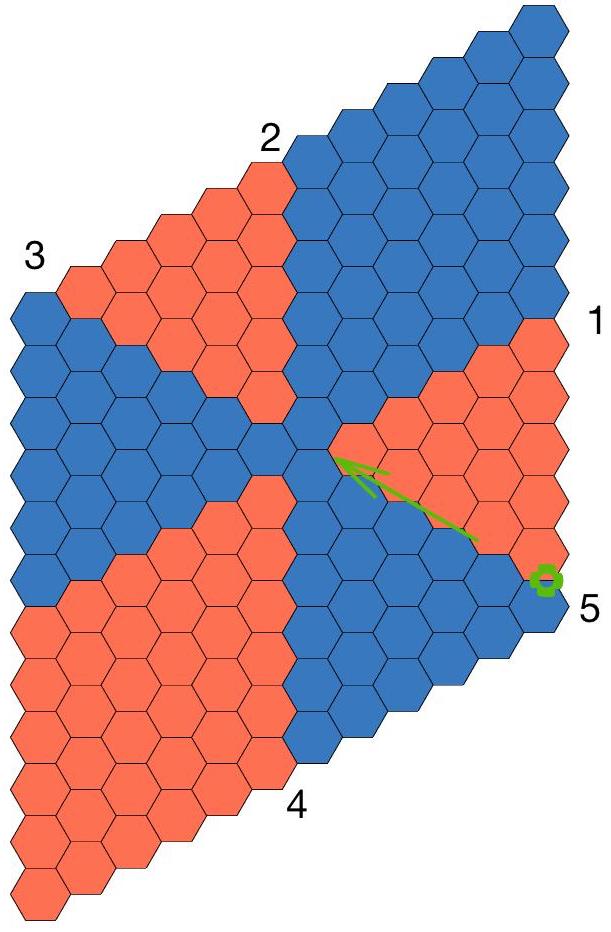}
\label{fig:Starfield_Schematic}
	\includegraphics[scale=0.155]{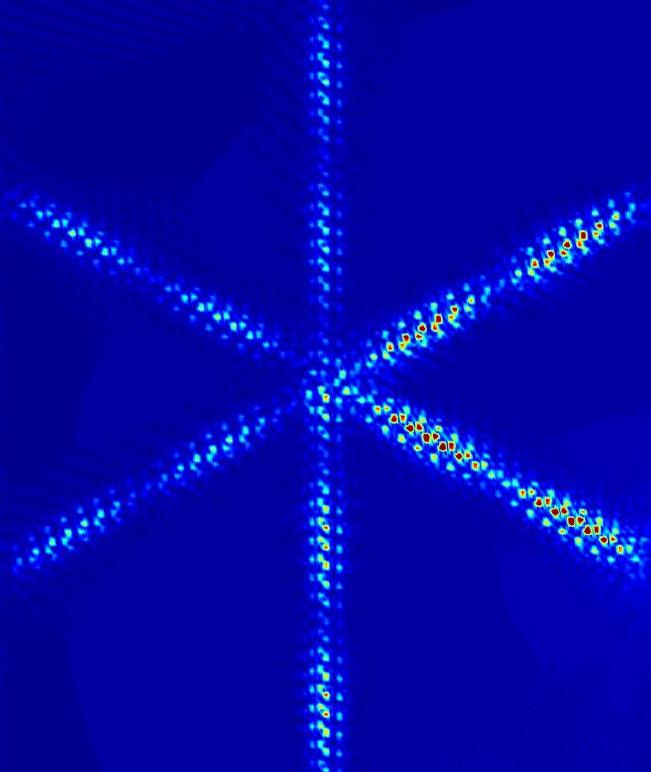}
\end{tabular} 
\caption{Figure illustrates $5$-way splitting of wave energy away from nodal region. Blue and orange cells associated with geometrically distinct cells (see Fig. \ref{fig:C3v_nonorthog_DC} for cellular structures).  Frequency $\omega = 18.85$; despite the bulk gap being $\{17.05, 19.90\}$, the 5-way splitter was only seen for a narrow band range of frequencies. This is in contrast to the more broadband 3-way energy-splitters, shown in Figs. \ref{fig:NatMat_Ex}, \ref{fig:NatComm_Ex}. Note the difference, in the nodal region, between this example and Fig. \ref{fig:Starfield_tot}.
}
\label{fig:five_splitter}
\end{figure}

Motivated by \cite{cheng_acoustic_2016, xiaoxiao_direct_2017}, that showed 2-way energy-splitters, we demonstrate how the $C_{3v}$ trivial case allows for enhanced 3-way splitting of energy for the same arrangement of distinct media as \cite{cheng_acoustic_2016, xiaoxiao_direct_2017} (Figs. \ref{fig:NatMat_Ex}, \ref{fig:NatComm_Ex}). After these comparative set of examples, we push those designs further, by concluding this subsection, with a novel 5-way energy-splitter, Fig. \ref{fig:five_splitter}.

Ref. \cite{xiaoxiao_direct_2017} demonstrated an `X' shaped design for a topological energy-splitter, Fig. \ref{fig:NatComm_Ex}(a). Their results indicated that the highest transmission occurred along ZLM $4$, followed closely by ZLM $2$; in contrast, to these two leads, negligible transmission occured along lead $3$. These results can be interpreted through the lens of the design paradigm (Sec. \ref{sec:network_paradigm}) and the use of the $C_{6v}$ nontrivial case. From Fig. \ref{fig:NatComm_Ex}(b), the distinctive modal shapes, and hence the different interfaces, are clearly evident (Sec. \ref{sec:modal}). The relatively higher transmission along lead $4$ compared with $2$ can be attributed to the design of the nodal region (Sec. \ref{sec:nodal}); when the wavelength of the incoming ZLM is comparable to the cell, then the two orange cells within the nodal region, invariably forms an effective barrier which limits propagation along lead $2$. The absence of propagation along lead $3$, as noted in \cite{xiaoxiao_direct_2017}, is due to the mismatch in phase of the mode which has an identical chirality to ZLM $1$ (Sec. \ref{sec:chirality_phase}). The absence of phase matching, between leads $1$ and $3$, can be rectified by replacing the $C_{6v}$ topologically distinct regions with $C_{3v}$ topologically nontrivial (albeit geometrically distinct) regions. The resulting scattered field, Fig. \ref{fig:NatComm_Ex}(c), shows propagation along lead $3$ and collectively, $3$-way energy-splitting, away from the nodal region. Despite the topological charge not being conserved \cite{ezawa_topological_2013}, the Fourier space separation (Sec. \ref{sec:Fourier_space}) between modes of opposite group velocity ensures a degree of robustness.

\onecolumngrid

\twocolumngrid

A similar example, to Fig. \ref{fig:NatComm_Ex}, is the topological network examined in \cite{cheng_acoustic_2016}; Fig. \ref{fig:NatMat_Ex}(c) imitates their example using the $C_{6v}$ nontrivial case. The relative transmission along active leads, difference in modal shape and absence of propagation along lead $1$ (Fig. \ref{fig:NatMat_Ex}) are all explained in a similar manner to the earlier example. The trivial analogue, Fig. \ref{fig:NatMat_Ex}(d), leverages the phase and group velocity matching $C_{3v}$ counterexample shown in Sec. \ref{sec:chirality_phase}, to allow for novel 3-way energy-splitting.


An example, that is independent of any pre-existing tessellations within the valleytronics literature, is the $5$-way splitter shown in Fig. \ref{fig:five_splitter}. Recall that we restrict ourselves to zigzag interfaces because they afford the maximum Fourier separation (see Sec. \ref{sec:Fourier_space}) between opposite propagating modes; hence our tessellation can comprise of, at most, 6 geometrically distinct regions. 
Therefore, the novel $C_{3v}$ trivial network, Fig. \ref{fig:five_splitter}, partitions energy, away from the nodal region, the maximum number of ways possible given the zigzag interface constraint.

\subsection{Illustrating 4-way splitting via tunneling using $C_{6v}$ nontrivial modes}
\label{sec:tunnel_network}

 \begin{figure} [h!]
\begin{tabular}{ll} 
	\includegraphics[scale=0.090]{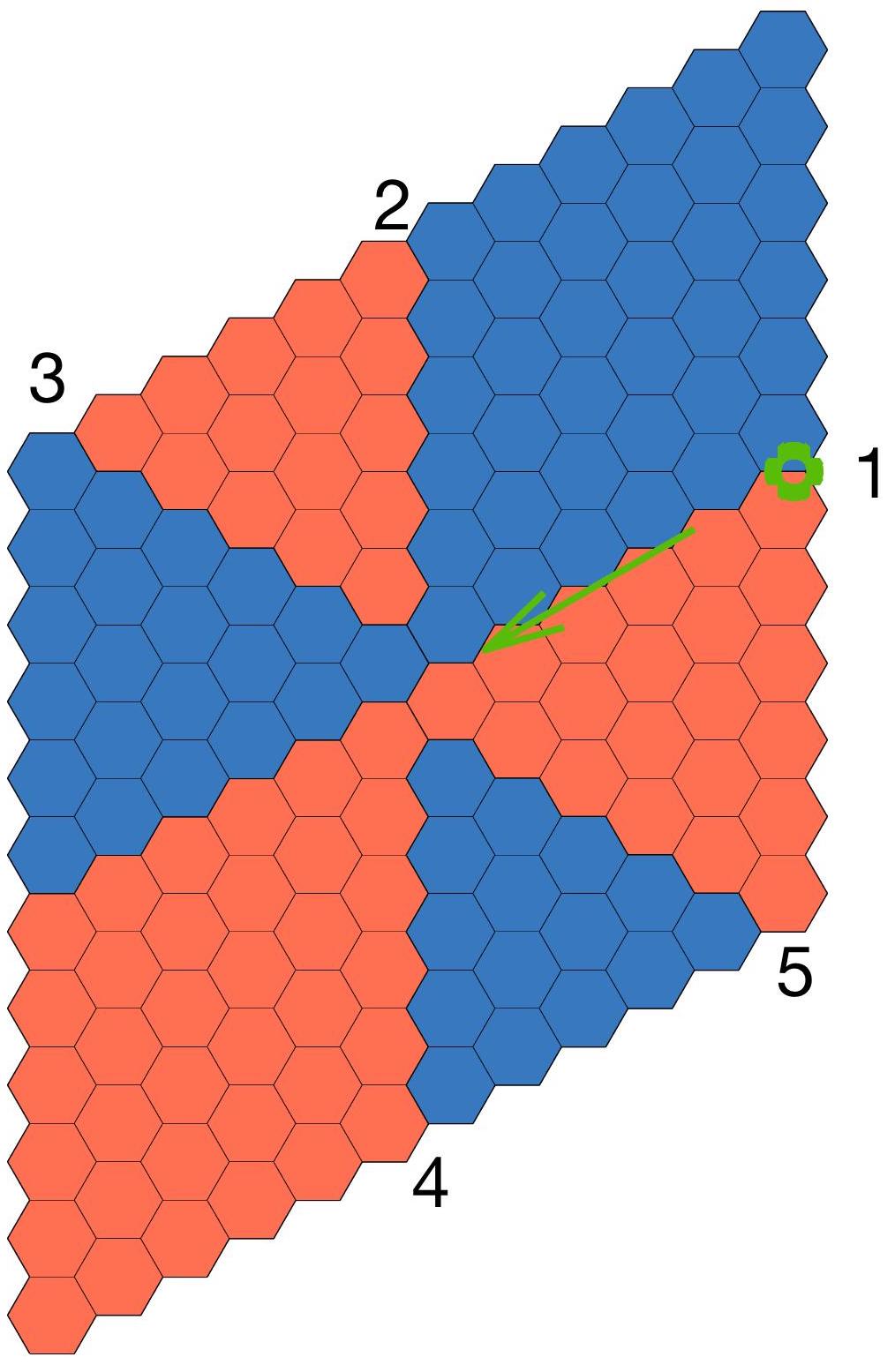}
	\includegraphics[scale=0.245]{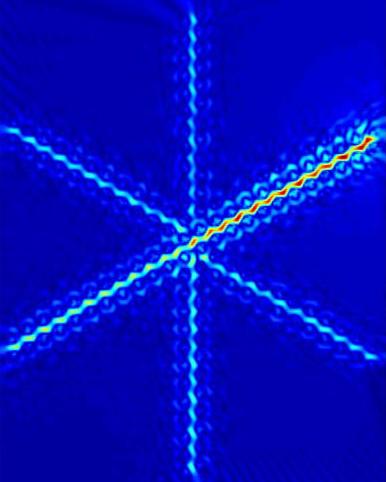}
\label{fig:Starfield}
\end{tabular} 
\caption{Figure illustrates $4$-way splitting of wave energy away from nodal region,  energy couples from
one topological valley mode into four others via tunneling. Blue and orange cells associated with opposing Chern valley numbers at a specific valley.  Frequency $\omega = 15.91$, bulk gap $\{13.90, 16.23\}$}
\label{fig:Starfield_tot}
\end{figure}

The novel topological network exemplar in this article, that contains a more than $2$-way energy-splitter, is shown in Fig. \ref{fig:Starfield_tot}. The tesselation is comprised of $4$ (at first sight $6$, but note the detail of the central nodal region) geometrically distinct regions, formed from the $C_{6v}$ nontrivial case, and containing a $4$-way
energy-splitter. The excited ZLM $1$ propagates through the junction,
and continues along the lead beyond it, however energy is shifted to ZLMs $2-5$. 
This $4$-way energy-splitting arises due to two sets of
tunneling (Sec. \ref{sec:tunneling}) occurring before and after the junction. The nodal region differs from 
Fig. \ref{fig:five_splitter}; this is done to ensure the propagation of ZLM $1$ 
through the central region and beyond. If instead we used the tessellation of Fig. \ref{fig:five_splitter},
this would create an effective barrier at the junction,
consequently we would obtain a similar arrow modal pattern to Fig. \ref{fig:filter_network}(b), albeit with backscattering.
A major benefit of utilising tunneling to partition energy is that it allows for the preservation of topological charge \cite{ezawa_topological_2013} and hence ZLMs $2-5$ in Fig. \ref{fig:Starfield_tot} are topologically protected; therefore, compared with the trivial energy-splitters, (Figs. \ref{fig:NatMat_Ex}(d), \ref{fig:NatComm_Ex}(c) and \ref{fig:five_splitter}) the nontrivial modes (Fig. \ref{fig:Starfield_tot}) are more robust and hence of more practical use.

\subsection{Topological supernetwork}
\label{sec:supernetwork}

\begin{figure} [htb!]
\centering
\begin{tabular}{ll}
(a) Supernetwork Tesselation & (b) Topological Supernetwork \\

		\includegraphics[scale=0.170]{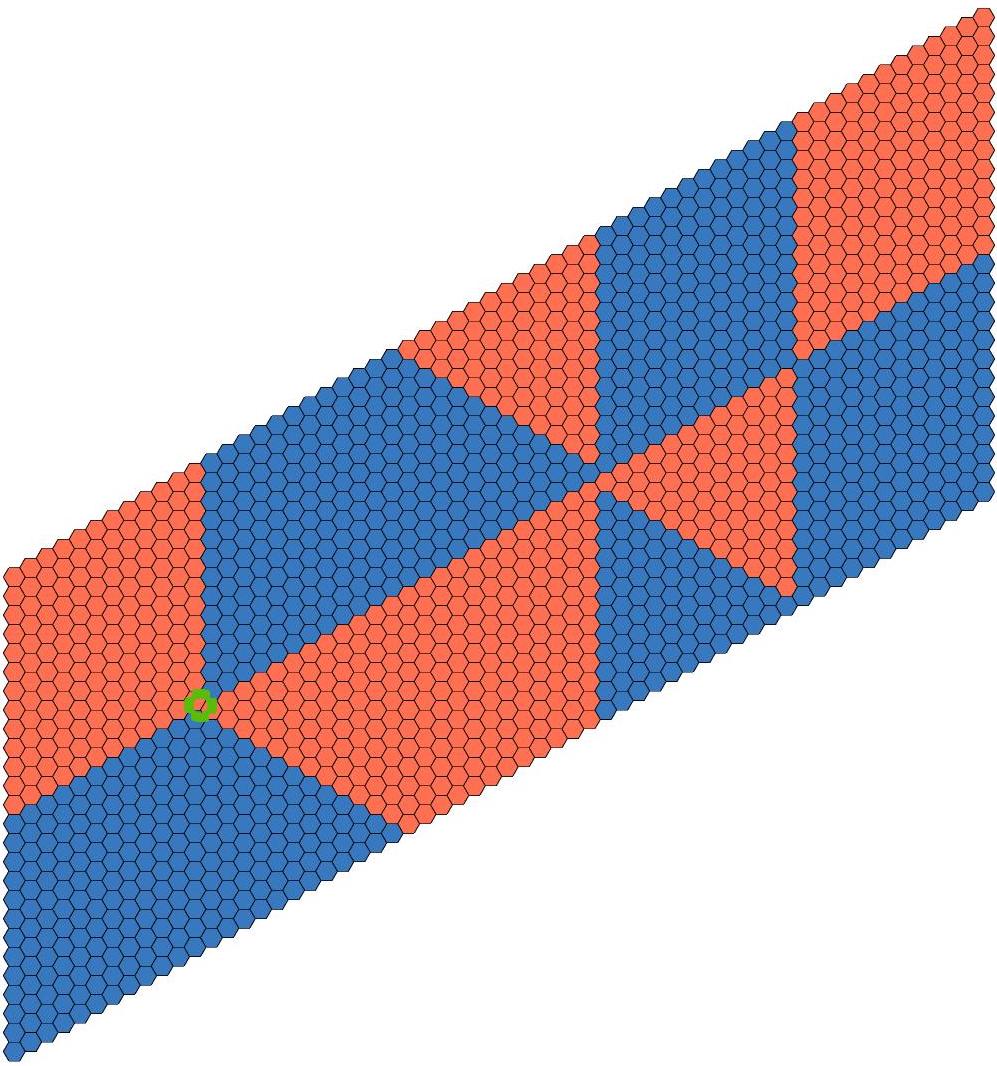}
\label{fig:supernetwork_tess}
&		\includegraphics[scale=0.315]{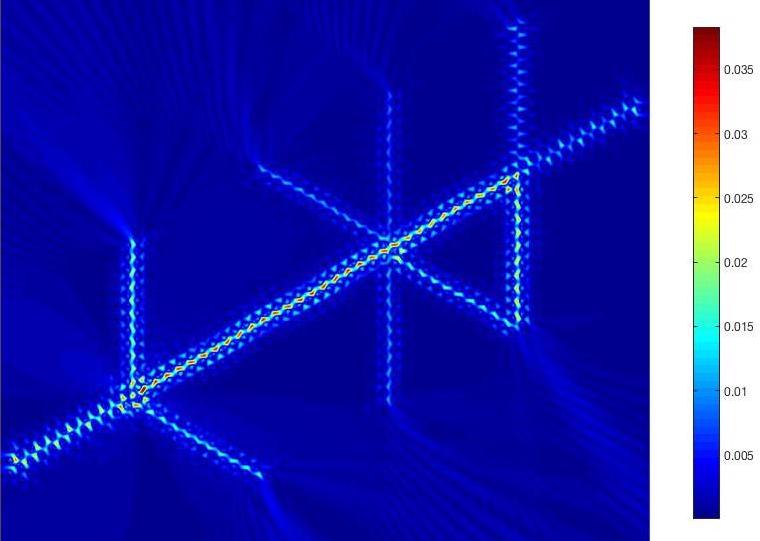}
		\label{fig:supernetwork_field}
		\end{tabular}
\caption{The total
                  arrangement contains $2340$ cells, each contains a
                  hexagonal arrangement of point scatterers; different colours denote dissimilar 
arrangements. The network is
                  excited at the leftmost junction with a dipole
				at $\omega = 15.28$. The colour bar is
                                for a linear gradient of values.}
\label{fig:supernetwork}
\end{figure}

The  topological supernetwork, 
Fig. \ref{fig:supernetwork}, 
 is generated using the design paradigm building blocks, Sec. \ref{sec:network_paradigm}. It contains only the $C_{6v}$ nontrivial case and therefore the topological charge is conserved; the giant tessellation is a combination of those tessellations found in Figs. \ref{fig:filter_network}(c), \ref{fig:Starfield_tot}, \ref{fig:NatComm_Ex}(b) in that order.
 
This is just one amongst many supernetworks that can now be
 constructed from individual building blocks. There are a myriad of other complicated topological networks that can now be accurately
 designed by partitioning the medium differently. Direct
 generalisations include, using fractional cells, different edge
 terminations or combining different geometrical cases. For the
 latter, we could design the $C_{6v}$ and $C_{3v}$ nontrivial cases
 to have simultaneous bulk bandgaps, then create a tesselation where neighbouring regions are 
topologically distinct (opposite valley Chern numbers). 
This way the filtering properties of the $C_{3v}$ case are combined with the 
dual propagation properties of the $C_{6v}$. Moreover, one can tune the rates of
decay of the ZLMs perpendicular to the interface; thereby controlling the amount of energy partitioned via tunneling as well as the  sensitivity of modes  
to backscattering by defects. 

\onecolumngrid

\twocolumngrid

\section{Concluding Remarks}
\label{sec:concluding}
Herein we have shown in detail how to design novel geometrically engineered interfacial wave networks  containing energy-splitters that partition energy in more than $2$-way directions. The main concepts used to design these systems have been laid out systematically in Sec. \ref{sec:network_paradigm}.
The specific model we use, the elastic plate and point masses, is irrelevant to our main argument which relies on topology and 
group theoretic principles. Thus we anticipate that the approach described will motivate the design of experimental, and other theoretical,
topological networks for all similar scalar wave systems: plasmonics, photonics, acoustics, as well as, for vectorial systems
such as plane-strain elasticity, surface acoustic waves and
 Maxwell equation systems). It is also easy to construct
geometries that do not involve point scatterers, see
Fig. \ref{fig:other_cells}, and yet will share the same group and geometric properties required for our designs. Additionally, returning to the
valley-Hall ``weak" topological phase that underlies the ZLMs: The principles that underly the topological networks presented here, should extend to potentially more robust geometrically induced phases
\cite{wu_scheme_2015, chang_multiple_2017, xiao_synthetic_2015}, thereby bringing the design of broadband, robust,
energy-splitters to yet another level.

Both authors thank the EPSRC for their support through grant
{EP/L024926/1} and R.V.C acknowledges the support of a Leverhulme Trust
Research Fellowship.

\bibliographystyle{apsrev4-1}
%

\end{document}